\documentclass[%
 reprint,
 amsmath,amssymb,
 aps,
prb,
]{revtex4-2}

\usepackage{graphicx}
\usepackage{dcolumn}
\usepackage{bm}
\usepackage{braket}
\usepackage{hyperref}

\usepackage{xcolor}
\usepackage{dsfont}

\newcommand{\beginsupplement}{%
	\setcounter{table}{0}
	\renewcommand{\thetable}{S\arabic{table}}%
	\setcounter{figure}{0}
	\renewcommand{\thefigure}{S\arabic{figure}}%
	\renewcommand{\theequation}{S\arabic{equation}}
}

\begin{document}

\title{Non-Abelian Anyons in Periodically Driven Abelian Spin Liquids}%

\author{Francesco Petiziol}
 \email{f.petiziol@tu-berlin.de}
 \affiliation{Technische Universit{\"a}t Berlin, Institut f\"ur Theoretische Physik, Hardenbergstra\ss e 36, 10623 Berlin}

\begin{abstract}
We show that non-Abelian anyons can emerge from an Abelian topologically-ordered system subject to local time-periodic driving. This is illustrated with the toric-code model, as the canonical representative of a broad class of Abelian topological spin liquids. The Abelian anyons in the toric code include fermionic and bosonic quasiparticle excitations which see each other as $\pi$ fluxes, namely they result in the accumulation of a $\pi$ phase if wound around each other. Non-Abelian behaviour emerges because the Floquet modulation can engineer a non-trivial band topology for the fermions, inducing their fractionalization into Floquet-Majorana modes bound to the bosons. The latter then develop non-Abelian character akin to vortices in topological superconductors, realizing Ising topological order. Our findings shed light on the nonequilibrium physics of driven topologically-ordered quantum matter and may facilitate the observation of non-Abelian behaviour in engineered quantum systems.
\end{abstract}

\maketitle

Quasiparticles with anyon quantum statistics are predicted to appear as gapped excitations in topologically-ordered (TO) quantum phases of matter, such as fractional quantum Hall and spin liquids~\cite{Nayak2008, Savary2017}. Braiding anyons of Abelian type can change the many-body wavefunction by a phase factor. Braiding non-Abelian anyons, instead, can induce a unitary transformation in a topologically-protected degenerate ground state, enabling fault-tolerant quantum computation~\cite{Kitaev2003, Kitaev2006, Nayak2008}. Abelian anyons have been recently observed and manipulated in various engineered quantum systems, ranging from ultracold atoms in optical lattices~\cite{Dai2017, Kwan2023} to Rydberg-atom arrays~\cite{Semeghini2021} and superconducting circuits~\cite{Satzinger2021, Andersen2023}, and first realizations of non-Abelian TO states have been reported in trapped ions quantum processors~\cite{Iqbal2024}. This progress motivates the study of TO phases out of equilibrium, such as subject to external control, which may both provide a stepping stone towards non-Abelian physics~\cite{Kitaev2006, Bombin2010, Wootton2008, Song2019, Andersen2023} and unveil novel effects, with examples being radical chiral Floquet phases~\cite{Po2016, Po2017} and fractionalized prethermalization~\cite{Jin2023}.

Here we show that time-periodic driving can effectively turn an Abelian TO system into a non-Abelian one. We use as a testbed the paradigmatic toric-code model~\cite{Kitaev2003}, which is believed to capture the fundamental low-energy physics of a large class of Abelian topological spin liquids~\cite{Savary2017, Sachdev2023} while remarkably being exactly soluble. This argues for the broad applicability of our results while enabling a clear illustration of the desired effect through exact methods. Abelian anyons in $\mathds{Z}_2$ spin liquids like the toric code emerge as bosonic and fermionic quasiparticle excitations which see each other as $\pi$ fluxes: Winding a fermion around a boson (and vice versa) results, unusually, in the accumulation of a $\pi$ phase. We show that a local Floquet modulation can alter the picture by driving one anyon type, the fermion, into a superconducting phase with a non-trivial band topology. This induces the fractionalization of the fermions into pairs of Majorana modes which bind to the other anyon type (the boson), realizing non-Abelian Ising TO~\cite{Nayak2008, Kitaev2006, Moore1991}. This is predicted through analytical high-frequency expansions and verified by numerical computation of the non-Abelian exchange phases. 
Our results shed light on the properties of driven interacting topological systems and promote the development of protocols to Floquet-engineer non-Abelian anyons~\cite{Kalinowski2023, Sun2023} from Abelian phases, further motivated by the recent experimental realizations~\cite{Semeghini2021, Satzinger2021, Bluvstein2022} and theoretical proposals~\cite{Weimer2010, Sameti2017, Verresen2021, Petiziol2024} of toric-code TO in quantum simulators. 
 Complementary to achievements in the manipulation of individual TO states~\cite{Satzinger2021, Andersen2023, Iqbal2024}, a Floquet-engineering approach has the appeal of realizing the background Hamiltonian, which stabilizes the desired TO states, makes anyons well-defined quasiparticles~\cite{Jiang2008} and enables the study of dynamics.

\begin{figure}[t]
\includegraphics[width=\linewidth]{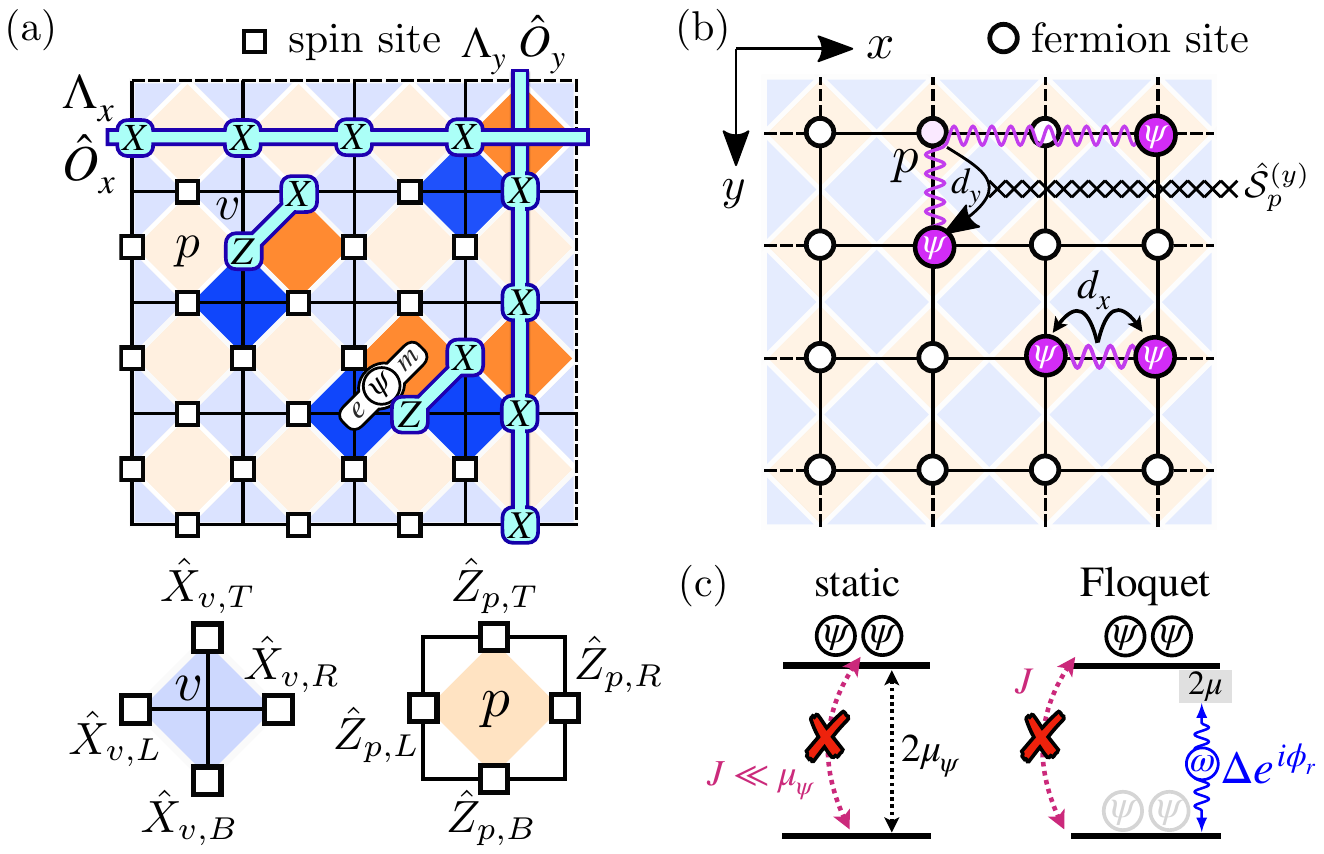}
\caption{(a) Toric-code spin lattice. The $m$ anyons live on plaquettes $p$ (orange), while $e$ anyons live on vertices $v$ (blue), and their bound state forms a fermion $\psi$. The operators $\hat{X}_i\hat{Z}_j$ depicted induce fermion tunnelling and pairing. Dashed lines indicate periodic boundaries. (b) Corresponding lattice for $\psi$ fermions in the quasiparticle mapping, where $d_x$ and $d_y$ control fermion tunnelling and pairing. Vertical processes include a boson operator $\hat{\mathcal{S}}_p^{(y)}$ reproducing the mutual statistics of $\psi$ and $e$. (c) Drive-assisted fermion pairing, restoring otherwise off-resonant ($J\ll \mu_\psi$) processes with rate $\Delta$ and phase $\phi_r$.}
\label{fig1}
\end{figure}

\textit{The driven model.---} The toric-code model~\cite{Kitaev2003, Wen2003} describes spin-1/2 systems on the bonds of a square $L\times L$ lattice [Fig.~\ref{fig1}(a)] ($L$ even hereafter) with Hamiltonian
\begin{equation} \label{eq:tcH}
\hat{H}_{\mathrm{tc}} = - \frac{g}{2} \sum_{v }\hat{A}_{v}\ - \frac{g}{2} \sum_{p} \hat{B}_{p} \ .
\end{equation}
The labels $p$ and $v$ denote lattice plaquettes and vertices, $g$ is a coupling constant, $\hat{A}_v=\prod_{i\in v} \hat{X}_i$ and  $\hat{B}_p=\prod_{i\in p} \hat{Z}_i$ are four-spin interactions and $\hat{X}_i, \hat{Y}_i, \hat{Z}_i$ are Pauli matrices for the $i$th spin. When convenient, we will also interpret $p=(p_x, p_y)$ and $v=(v_x,v_y)$ as coordinates of two separate plaquette and vertex lattices, respectively, both with lattice vectors $x=(1,0)$ and $y=(0,1)$. All four-spin operators in~\eqref{eq:tcH} commute with each other, making the model exactly solvable. The eigenstates can be labelled by the eigenvalues $A_v=\pm 1$ and $B_p=\pm1$ for each vertex and plaquette. On a torus, since $\prod_v \hat{A}_v = \prod_p \hat{B}_p=\mathds{1}$, a complete set of observables is obtained by including two operators $\hat{O}_x=\prod_{i\in\Lambda_x}{\hat{X}_i}$ and $\hat{O}_y=\prod_{i\in \Lambda_y} {\hat{X}_i}$ spanning noncontractible loops [$\Lambda_x$ and $\Lambda_y$ in Fig.~\ref{fig1}(a)] and commuting with all $\hat{A}_v$ and $\hat{B}_p$. The $\pm1$ eigenvalues of $\hat{O}_{x/y}$ define four superselection sectors, leading to four degenerate ground states satisfying $ A_v= B_p = 1$. Flipping one eigenvalue, $A_v=-1$ ($ B_p = -1$), costs an excitation energy $g$ and is interpreted as the creation of a pointlike quasiparticle $e$ ($m$) on top of the ground state $\mathds{1}_{\rm tc}$, which occupies the corresponding vertex $v$ (plaquette $p$). 
 While $e$ and $m$ are hardcore bosons under self-exchange, they are mutual semions: the wavefunction picks up a Peierls-like $\pi$ phase when $m$ is wound around $e$, or vice versa. A bound state $\psi =e\times m$, costing an energy $\mu_\psi=2g$, is then a fermion, while still being a semion with respect to a separate $e$ or $m$. These quasiparticles, including $\mathds{1}_{\rm tc}$, represent the four anyon types in the toric code. They are Abelian, since their non-trivial fusion has a single possible outcome: $e\times m = \psi$, $e\times \psi = m$, $m \times \psi= e.$
 
We add a drive to $\hat{H}_{\rm tc}$ to induce $\psi$-fermion motion and realize a topological band structure. Non-Abelian character will emerge from the interplay between the fermion band topology, achieved through Floquet engineering, and the semion relation between the $\psi$ and $e$ Abelian anyons, provided by the underlying toric code. The drive addresses neighbouring spins as 
\begin{equation} \label{eq:driveH}
\hat{H}_d(t) = -\sum_{p} \big[ d_x(t)\hat{X}_{p,R} \hat{Z}_{p,B}  + d_y(t)  \hat{X}_{p,T} \hat{Z}_{p,L}\big].
\end{equation}
The subscripts $({p,j})$, with $j\in \{B,R,T,L\}$ denote the spins at the bottom, right, top or left of $p$, respectively [Fig.~\ref{fig1}(a)]. 
We expect drives of even simpler structure to achieve the desired effect, but \eqref{eq:driveH} has the key merit that, while still being local, it allows for a transparent analytical treatment and efficient numerical simulation of the time-dependent system, as we shall see, thus enabling a clear illustration of the target physics. This is an important advantage: investigating the toric code with even \textit{static} magnetic-field perturbations generically requires sophisticated methods not directly applicable to the driven case~\cite{Trebst2007, Tupitsyn2010, Dusuel2011}.
Two-spin couplings of the form $\hat{X}_i \hat{Z}_j$ on adjacent spins induce $\psi$ tunnelling and pairing: they displace a pair of neighbouring $e$ and $m$ composing $\psi$, as well as create and annihilate pairs of them. Indeed, since $\hat{X}_i$ anticommutes with $\hat{B}_p$ on neighbouring plaquettes $p$ and $p'$ while commuting with $\hat{A}_v$, its action on an eigenstate flips the eigenvalues $B_p$ and $B_{p'}$, while preserving the joint $m$-anyon parity $(-1)^{B_p+B_{p'}}$. Hence, if a $m$-particle composing a fermion is present at $p$, it tunnels to $p'$ (and vice versa). If no fermion or a pair is present, the associated pair of $m$ will be created or annihilated. An equivalent analysis applies to $e$-motion on nearby vertices induced by $\hat{Z}_j$. 
The functions $d_x(t)$ and $d_y(t)$ control processes along the horizontal and vertical direction, respectively. 

\textit{Quasiparticle picture.---} With the chosen form of the modulation, the driven model can be mapped exactly to a problem of driven noninteracting $\psi$ fermions coupled to static $e$ bosons, which allows us to focus on the impact of the drive on the $\psi$ anyons. To this end, we adopt the quasiparticle picture of Ref.~\cite{Chen2022} (details are given in the Supplemental Material (SM)~\cite{SM}). The spins' Hilbert space is mapped to the tensor product of Fock spaces for $e$ hardcore-boson and $\psi$ fermion occupations, and of a four-dimensional Hilbert space reflecting four superselection sectors. The spin operators map to creation and annihilation operators of bosons on vertices ($\hat{b}_v^\dagger, \hat{b}_v$) and of fermions on plaquettes ($\hat{f}_p^\dagger, \hat{f}_p$), with the toric-code ground state $\ket{0_{\rm tc}}$ representing the quasiparticle vacuum. The Hamiltonian $\hat{H}_{\rm tc}$ maps to $\hat{H}_e+\hat{H}_{e\psi}^{(0)}$ with $\hat{H}_e= g \sum_v \hat{b}_v^\dagger \hat{b}_{v}$ and $\hat{H}_{e\psi}^{(0)}=2g\sum_p [1-\hat{b}^\dagger_{v(p)}\hat{b}_{v(p)}]\hat{f}_p^\dagger \hat{f}_p$, describing quasiparticles at rest [$v(p)$ denotes the vertex to the bottom-left of $p$]. 
The drive $\hat{H}_d(t)$ maps to $\hat{H}_{e\psi}^{(d)}(t)= - \sum_{p,r\in\{x, y\}}  d_{r}(t) \hat{\mathcal{S}}_{p}^{(r)}  (\hat{f}_{p} \hat{f}_{p+r} 
+ \hat{f}_{p}^\dagger \hat{f}_{p+r} + \mathrm{H. c.})$, describing fermion tunnelling and pairing but also containing a coupling to the $e$-bosons through $\hat{\mathcal{S}}_{p}^{(x)}=1$ and $\hat{\mathcal{S}}_{p}^{(y)}=\prod_{v\in{\mathrm{R}[v(p)]}}(-1)^{\hat{b}_v^\dagger \hat{b}_v}$. Here, $\mathrm{R}[v(p)]$ denotes the vertices to the right of the spin shared by the plaquettes $p$ and $p+y$, see Fig.~\ref{fig1}(b). This coupling explicitly reproduces the semion mutual statistics: $\hat{\mathcal{S}}_{p}^{(r)}$ yields the accumulation of $\pi$ phase when a fermion encircles a $e$ particle. Since the drive does not induce boson motion, all bosonic terms are diagonal in the Fock basis $\ket{\vec{n}^e}=(\hat{b}_{v_1}^\dagger)^{n_1}\dots (\hat{b}_{v_N}^\dagger)^{n_N}\ket{0_{\rm tc}}$ describing $e$ occupations. We can then analyze the fermion Hamiltonian $\hat{H}_\psi(t)= \bra{\vec{n}^e}\hat{H}_{e\psi}^{(0)}  + \hat{H}_{e\psi}^{(d)}(t)\ket{\vec{n}^e}$ corresponding to a fixed distribution of $e$ particles. The potential presence of $e$-bosons at vertices $\{v\}$ is assumed to derive from preparing the ground state of a slightly modified Hamiltonian~\eqref{eq:tcH} with inverted coupling $g\to\tilde{g}=-g$ for those vertices. The Hamiltonian $\hat{H}_\psi(t)$ then reads
\begin{multline}
\hat{H}_\psi(t)  =   \sum_{p} \Big[\mu_{\psi} \hat{f}_{p}^\dagger \hat{f}_{p}\\- \!\! \sum_{r\in\{x, y\}}  d_{r}(t) {\mathcal{S}}_{p}^{(r)}  \big(\hat{f}_{p} \hat{f}_{p+r} 
+ \hat{f}_{p}^\dagger \hat{f}_{p+r} + \mathrm{H. c.}\big) \Big]. 
\label{eq:Hmapped}
\end{multline} 
where $\mathcal{S}_p^{(r)} = \bra{\vec{n}^e}\hat{\mathcal{S}}_p^{(r)} \ket{\vec{n}^e}$. It describes spinless superconducting fermions coupled, through $\hat{\mathcal{S}}_{p}^{(r)}$, to an effective gauge flux given by a background distribution of $e$ particles. For a fixed $e$-boson distribution, the driven toric code then maps to a problem of driven \textit{noninteracting} fermions.

\textit{Floquet engineering.---} The above ingredients suggest a potential analogy to topological superconductors~\cite{Read2000, Nayak2008, Sato2017}, where non-Abelian anyons emerge as vortices ($\sigma$) carrying Majorana zero modes~\cite{Ivanov2001}. The latter result from the fractionalization of the fermions occurring when their bulk bands are gapped and topological, as is the case for pairing in so-called $p+ip$ symmetry~\cite{Nayak2008}. In the lattice model~\eqref{eq:Hmapped}, this corresponds to a complex-valued pairing coupling $\Delta_{r} f_{p}^\dagger f_{p+r}^\dagger $ with $(\Delta_{x},\Delta_{y})=(\Delta, i\Delta)$~\cite{Sachdev2023}. These so-called Ising anyons $\sigma$, $\mathds{1}_{\rm is}$ and $\psi$ feature non-trivial fusion rules $\sigma \times \sigma = \mathds{1}_{\rm is} + \psi$, $\psi \times \psi = \mathds{1}_{\rm is}$, $\psi\times \sigma=\sigma$~\cite{Nayak2008, Kitaev2006}. The multiple fusion channels for the vortex $\sigma$, yielding either the vacuum $\mathds{1}_{\rm is}$ or a fermion $\psi$, qualify them as non-Abelian. We will show that the driven toric code reproduces faithfully this physics, with $e$ particles behaving like $\sigma$ vortices, for appropriate driving functions that Floquet-engineer complex fermion pairing. 

Since the drive controls the physical spins, rather than the quasiparticle pairing terms directly, no choice of time-\textit{in}dependent $d_x$ and $d_y$ in Eq.~\eqref{eq:driveH} can achieve the goal. Indeed, the functions $d_{r}(t)$ need be real-valued and they induce tunnelling and pairing with equal real rate. We overcome these limitations employing a periodic modulation $d_{r}(t)=d_{r}(t+T)$ and Floquet engineering~\cite{Bukov2015, Eckardt2017}. The driving functions are chosen as
\begin{equation} \label{eq:drivef}
d_{r}(t) = J + 2 \Delta \cos(\omega t + \phi_{r}),
\end{equation}
with frequency $\omega=2\pi/T$ and amplitudes $J$ and $\Delta$ much smaller than the fermion chemical potential, $J,\Delta\ll \mu_{\psi}$. They only differ in their phase $\phi_{r}$.  
Consider first the limit $\Delta=0$ in Eq.~\eqref{eq:drivef}. Since the energy $2\mu_\psi$ required for pair creation and annihilation is much larger than $J$, pairing processes with a real coupling $J$ in~\eqref{eq:Hmapped} are far off-resonant and effectively suppressed, leaving only quasiparticle tunnelling. They are restored via the modulation with $\Delta\ne 0$ and a frequency $\omega$ quasiresonant with $2\mu_\psi$ [Fig.~\ref{fig1}(c)]. The advantage is that a complex phase can be attributed to the `photon-assisted' pairing coupling, depending on the phases $\phi_{r}$ of the drives. We choose $\omega=2(\mu_{\psi}+\mu)$, allowing for a small detuning $2\mu$. Applying Floquet theory~\cite{Shirley1965, Sambe1973}, the stroboscopic dynamics induced by the time-periodic Hamiltonian~\eqref{eq:Hmapped} in steps of $T$ is captured as $\hat U(nT)=e^{-i\hat{H}_FnT}$ by a time-independent Floquet Hamiltonian $\hat{H}_{F}$, which can be approximated from $\hat{H}_\psi(t)$ through high-frequency expansions, in the regime $\omega\gg J,\Delta,\mu$ considered here~\cite{Goldman2014, Bukov2015, Eckardt2015, Mikami2016, Eckardt2017}. To leading order in $\omega^{-1}$ and in a frame defined by $\hat{R}(t)=\exp[-i t(\omega/2)\sum_{p} \hat{f}_{p}^\dagger \hat{f}_{p}]$~\cite{SM}, $\hat{H}_F$ is approximated by the time-averaged Hamiltonian 
\begin{align} 
\hat{H}_{\rm avg}  & = - \sum_{p}\Big[\mu \hat{f}_{p}^\dagger \hat{f}_{p} \nonumber \\ + & \sum_{r=x,y}  {\mathcal{S}}_{p}^{(r)} \ \big(\Delta e^{i\phi_{r}}  \hat{f}_{p} \hat{f}_{p+r} + J \hat{f}_{p}^\dagger \hat{f}_{p+r} + \mathrm{H. c.})\Big] \ .
\label{eq:HF}
\end{align}
At time $T$, $\hat{R}(T) = (-1)^{\sum_{p} \hat{f}^\dagger_{p}\hat{f}_{p}}$ reduces to the total fermion parity, which is a conserved quantity. 
Since quasiparticles in the toric code can only be created in pairs, only even-parity states are physical and $\hat{R}(T)=1$ in their subspace. 
The desired $p+ip$ pairing in $\hat{H}_{\rm avg}$ is obtained by choosing a circular-like shaking, with phase delay $\phi_{y} - \phi_{x} = \pi/2$ between horizontal and vertical modulations. 
The effective chemical potential $\mu$ is controlled by the detuning of the drive from the excitation energy of a fermion pair in the toric code, and the fermions are coupled to the $e$ particles via ${\mathcal{S}}_{p}^{(r)}$. 

 \begin{figure}[t]
\includegraphics[width=\linewidth]{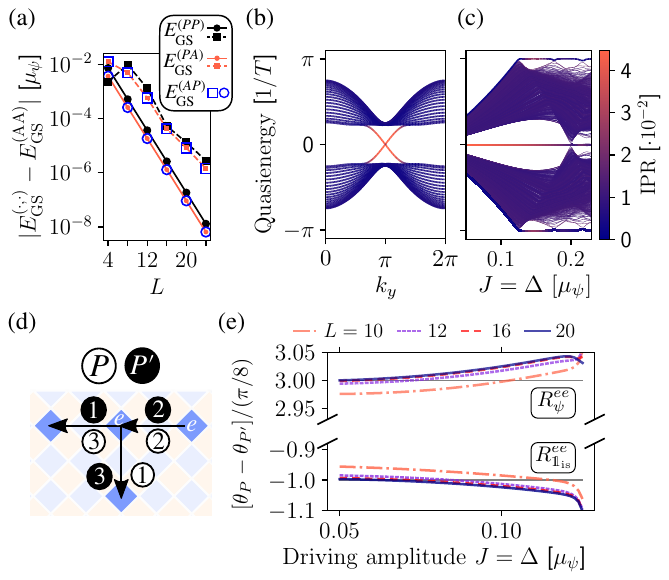}
\caption{(a) Splitting between fermion ground states for $\mu=-2J$ in different superselection sectors, having energy $E_{\rm GS}^{(x,y)}$ [$(x,y)$ denotes the boundary conditions, either periodic ($P$) or antiperiodic ($A$) along $x$ and $y$] in the high-frequency regime ($J=\Delta=0.1\mu_\psi$, solid lines and circles) and in the anomalous phase ($J=\Delta=0.15\mu_\psi$, dashed lines and squares). (b) Floquet-BdG quasienergies on a $L_x\times L_y=20\times200$ cylinder, for $J=\Delta=0.1\mu_\psi$, and (c) on a $(A,A)$-torus in the presence of a $e$ pair, for increasing driving amplitude ($\mu=-2J$, $L=24$). The color-scale indicates the inverse participation ratio of the corresponding eigenstate. (d) Paths $P$ and $P'$ involved in the vortex-exchange protocol. (e) Vortex exchange phase $\theta_P-\theta_{P'}$ in the $\mathds{1}_{\rm is}$ (bottom) and $\psi$ (top) sectors at $\omega=2(\mu_\psi-2J)$, for different driving amplitudes and system sizes.}
\label{fig2}
\end{figure}

\textit{Floquet-Majorana modes.---} We verify that this Floquet modulation indeed induces topological properties and yields Floquet-Majorana modes. We study these properties by numerically computing the Floquet Bogoliubov-de Gennes (BdG) Hamiltonian $\hat{H}_{F,\mathrm{BdG}}$, defined by $\hat{H}_F = (1/2)(\hat{\bm{f}}^\dagger, \hat{\bm{f}})\hat{H}_{F,\mathrm{BdG}}(\hat{\bm{f}}, \hat{\bm{f}}^\dagger)^T$, as well as its quasienergy spectrum and ground state parity, from the driven model of Eq.~\eqref{eq:Hmapped}~\cite{SM}. The topological phase is expected for $|\mu|<4J$ and is characterized by a non-zero Chern number of the negative-energy fermion band~\cite{Read2000, Sato2017, Sachdev2023}. Tuning the frequency $\omega$ to this topological region ($\mu\simeq -2 J$), in the absence of $e$ vortices [$\mathcal{S}_{p}^{(r)}=1$], we find four nearly-degenerate high-frequency `ground' states on a torus, one in each superselection sector corresponding to periodic or anti-periodic boundary conditions in $x$ and $y$ direction~\cite{Read2000}. The topological degeneracy is approached exponentially with increasing system size, as shown in Fig.~\ref{fig2}(a), and only three ground states have even fermion parity and are thus physical, signalling three Ising-anyon types $e, \mathds{1}_{\rm is},\psi$. On a cylinder, the quasienergy spectrum exhibits two edge modes [Fig.~\ref{fig2}(b)], indicating the development of nontrivial topology and a non-zero Chern number. 

Introducing a pair of separated $e$ particles on the torus, approximate Majorana modes at near-zero quasienergy appear [Fig.~\ref{fig2}(c)], which persist until $\omega$ is detuned out of the topological phase, as predicted by the time-averaged description of Eq.~\eqref{eq:HF}. For weak driving the gap remains open as the amplitude $J=\Delta$ increases, indicating that higher-order terms neglected in~\eqref{eq:HF} do not fundamentally alter the nature of the phase in this regime. Their impact can be suppressed exponentially with decreasing $J/\omega$~\cite{Abanin2015, Mori2016}. While generic periodically-driven systems are expected to reach ``infinite-temperature'' at long times due to energy absorption from the drive~\cite{Dalessio2014, Moessner2017}, our model may evade such heating given that it maps to an integrable fermion system (and static bosons)~\cite{Russomanno2012, Lazarides2014b}. We verify in~\cite{SM} that this is indeed the case at high frequency, also for large $e$-particle densities. Increasing the driving amplitudes beyond the high-frequency regime, while maintaining $\mu=-2J$, the fermions enter an anomalous Floquet phase~\cite{Jiang2011, Rudner2013, Eckardt2017}. Here the time-averaged description of Eq.~\eqref{eq:HF} is no longer valid, but the system still exhibits Majorana modes, at both quasienergy zero and $\pm \pi/T$, see Fig.~\ref{fig2}(c). %
This effect, only possible in driven systems~\cite{Oka2009, Kitagawa2010, Jiang2011, Liu2013, Tong2013, Asboth2014, Benito2014, Wintersperger2020, Jangjan2022}, has been proposed as a means to realize non-Abelian braiding and quantum computation in 1D Kitaev quantum wires~\cite{Bomantara2018, Bomantara2018b, Bauer2019, Bomantara2020}. 

\textit{Non-Abelian exchange phases.---} We analyze whether the exchange statistics of two $e$ particles (vortices) in the fermion topological phase is non-Abelian, as predicted in the presence of Majorana zero modes~\cite{Kitaev2006, Ivanov2001}. 
Two vortices are exchanged explicitly in different fusion sectors following Levin-Wen's protocol~\cite{Levin2003, Kitaev2006, Kawagoe2020}, sketched in Fig.~\ref{fig2}(d), and Ref.~\cite{Chen2022}. The protocol computes the difference in the Berry phase accumulated by moving vortices along two paths $P$ and $P'$, sharing the same set of positions but involving a different order of vortex moves. This guarantees that the resulting phase is only determined by the exchange statistics~\cite{Levin2003}. For each intermediate vortex configuration $\ket{\vec{n}_i^e}$ along the paths, we numerically determine the corresponding even-parity ground state $\ket{\Phi_i}$ of the Floquet-BdG Hamiltonian~\cite{SM}. The Berry phase accumulated along the path $P$ is computed from the sequence of ground states as $\theta_P = \mathrm{Arg}\prod_{(i,i+1)\in P} \bra{\Phi_{i+1}} \hat{Z}^{f}_{i+1,i}  \ket{\Phi_i}$. Here, $\hat{Z}^{f}_{i+1,i}$ is the fermionic part of the quasiparticle representation of the spin operator $\hat{Z}_{i}$ which converts the vortex configuration $\ket{\vec{n}^e_i}$ into $\ket{\vec{n}^e_{i+1}}$ by displacing one $e$ particle. Details about the evaluation of the matrix elements in $\theta_P$ are given in the SM~\cite{SM}. The prediction for Ising anyons is that the exchange phase $R_{\mathds{1}_{\rm is}}^{\sigma\sigma}$ in the $\mathds{1}_{\rm is}$ fusion sector (where the vortices fuse to the vacuum) and $R_{\psi}^{\sigma\sigma}$ in the $\psi$ sector (where they fuse to a fermion $\psi$) differ by $\pi C/2$ for Chern number $C$, namely $R_{\mathds{1}_{\rm is}}^{\sigma\sigma}=e^{-i\pi {C}/8}$ and $R_{\psi}^{\sigma\sigma}=e^{3i\pi C/8}$~\cite{Kitaev2006}--- a signature of their non-Abelian nature~\cite{Cheng2011}. The two sectors $\mathds{1}_{\rm is}$ and $\psi$ are selected by creating the $e$ pair on top of two different ground states, corresponding to doubly-antiperiodic and doubly-periodic fermion boundary conditions, respectively~\cite{Chen2022}\footnote{In the antiperiodic case, the even-parity ground state is the quasiparticle vacuum, such that the pair of vortices is created from and fuses to the vacuum. For the periodic case, the BdG ground state has odd parity and thus the physical even-parity ground state is the first-excited state containing one BdG fermion. The vortex pair thus fuses to this fermion.}. The counterclockwise exchange phases obtained are shown in Fig~\ref{fig2}(e), for varying driving amplitude in the high-frequency regime and for different system size. The results converge rapidly with the size towards the prediction for Ising TO with $C=1$. We have thus shown that the $e$ vortices carrying Floquet-Majorana modes, arising in the driven toric code in the high-frequency regime, behave like non-Abelian Ising anyons. The stability of the exchange phases for increasing driving amplitude in this regime confirms that higher-order corrections to the time averaged Hamiltonian do not disrupt the topological phase, as anticipated. While the direct $e$ exchange probed here is a smoking-gun signature at zero temperature, fermion fractionalization can be detected at finite temperature, e.g., in the temperature dependence of Raman scattering intensities~\cite{Nasu2016}.

{\it Conclusion.---} We have shown that time-periodic driving of an Abelian-anyon system can induce non-Abelian topological order, using Kitaev's toric code as the paradigmatic example of a large class of Abelian topological phases.
Our findings suggest a potential path towards non-Abelian anyons in synthetic quantum systems, where Abelian phases have been realized~\cite{Semeghini2021, Satzinger2021}. The model studied extends the range of potential candidates exhibiting 2D Floquet-Majorana physics beyond systems with intrinsic superconductivity or superfluidity~\cite{Yang2018,Zhang2021} and offers a flexible playground where key parameters such as quasiparticle motion and pairing processes can be independently controlled via the drive. The toric code, known to be closely related to quantum dimer models~\cite{Misguich2002, Buerschaper2014, Iqbal2014}, has been recently shown to describe dimer liquids of Rydberg excitations constrained by Rydberg blockade~\cite{Tarabunga2022, Verresen2022, Samajdar2023}, where Abelian spin-liquid behaviour has been observed~\cite{Semeghini2021}. This represents a promising setup to explore the ideas presented here, alongside adaptations of Floquet-engineering protocols for the toric code proposed in superconducting circuits~\cite{Sameti2017, Petiziol2024}. 

{\it Acknowledgments.---} The author is grateful to Andr{\'e} Eckardt, Nathan Harshman, Alexander Schnell, Isaac Tesfaye and Sandro Wimberger for helpful and inspiring discussions. The author also acknowledges funding from the Deutsche Forschungsgemeinschaft (DFG, German Research Foundation) via the Research Unit FOR 2414 - project number No. 277974659.

\bibliography{biblio}

\cleardoublepage
 \let\oldaddcontentsline\addcontentsline
\renewcommand{\addcontentsline}[3]{}

\let\addcontentsline\oldaddcontentsline

\beginsupplement
\setcounter{page}{1}
\onecolumngrid
\begin{center}
{\bf \large Supplemental Material to} \\
\vspace{0.2cm}
{\bf \large Non-Abelian Anyons in Periodically Driven Abelian Spin Liquids} \\
\vspace{0.4cm}

{ Francesco Petiziol }\\

{\itshape
Technische Universität Berlin, Institut für Theoretische Physik, Hardenbergstra{\ss}e 36, Berlin 10623, Germany}
\end{center}

\vspace{.5cm}

\twocolumngrid

\section{Quasiparticle mapping}
\label{app:mapping}

We review in this section the quasiparticle mapping of Ref.~\cite{Chen2022} and apply it to the driven toric-code model of Eq.~(1) and~(2) on a torus. 

We use the notation $p(v)$ to indicate the plaquette situated to the immediate top-right of a vertex $v$, and, conversely, $v(p)$ to denote the vertex to the bottom-left of the plaquette $p$.
The set of commuting plaquette and vertex operators $\hat{B}_p$ and $\hat A_{v}$ of the toric code, whose $\pm 1$ eigenvalues are associated with the presence or the absence of $e$ and $m$ quasiparticles, represents a complete set of observables on the system Hilbert space on a infinite lattice. For an $L\times L$ lattice on a torus, hosting $2L^2$ spins and $L^2$ vertices and plaquettes, they are under-complete because of the condition $\prod_p \hat{B}_p=\prod_v \hat{A}_v=\mathds{1}$: Specifying all their (independent) plaquette and vertex eigenvalues imposes $2^{(L^2-1)}\cdot 2^{(L^2-1)}$ constraints in the $2^{L^2}$ Hilbert, leaving two spin-1/2 degrees of freedom free corresponding to the topological degeneracy. An equivalent commuting set, sharing the same common eigenstates of the operators $\hat{A}_v$ and $\hat{B}_p$, is given by the operators $\hat{A}_v\hat{B}_{p(v)}$ and $\hat{B}_{p}$, whose $\pm 1$ eigenvalues can be associated with the presence of $e$ and $\psi$ quasiparticles instead. A fermion $\psi$ is then conventionally interpreted as located on the plaquette $p$ and composed of a pair of $m$ and $e$ anyons occupying the plaquette $p$ and vertex $v(p)$, respectively (as depicted in Fig.~1). The presence (absence) of a fermion is detected by the $-1$ ($+1$) eigenvalue of $\hat{B}_p$, thus counting the parity $(-1)^{n_p^\psi}$ of the number of fermions $n_p^\psi\in\{0,1\}$ potentially occupying the plaquette $p$. Similarly, the presence (absence) of an isolated $e$ boson is detected by the $-1$ ($+1$) eigenvalue of $\hat{A}_v\hat{B}_{p(v)}$. We thus refer to $\hat{B}_p$ as fermion parity operators and to $\hat{A}_v\hat{B}_{p(v)}$ as boson parity operators.

\begin{figure}[b]
\centering
\includegraphics[width=\linewidth]{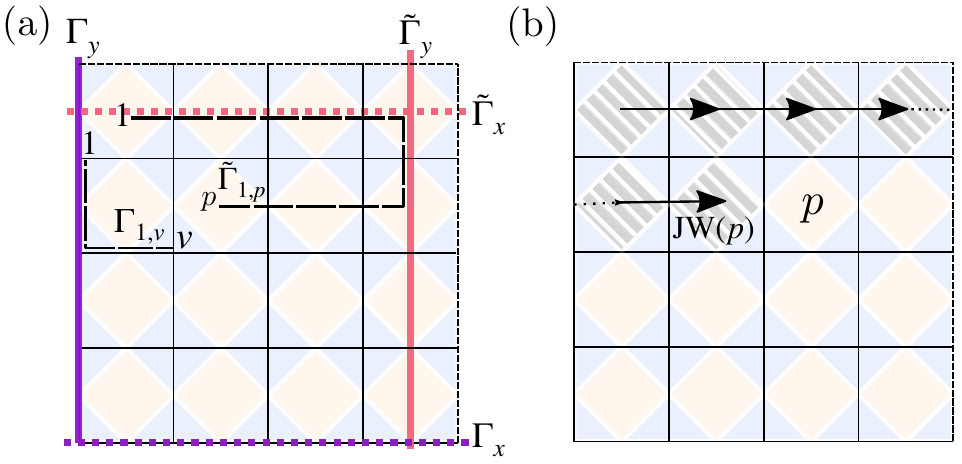}
\caption{Paths used in the quasiparticle mapping of Ref.~\cite{Chen2022} (a) for the pseudospin mapping ($\Gamma_{x/y}$, $\tilde{\Gamma}_{x/y}$, $\Gamma_{1,v}$ and $\tilde\Gamma_{1,p}$) and (b) in the Jordan-Wigner (JW) mapping to fermions.}
\label{fig:s1}
\end{figure}

\begin{figure}[t]
\centering
\includegraphics[width=0.9\linewidth]{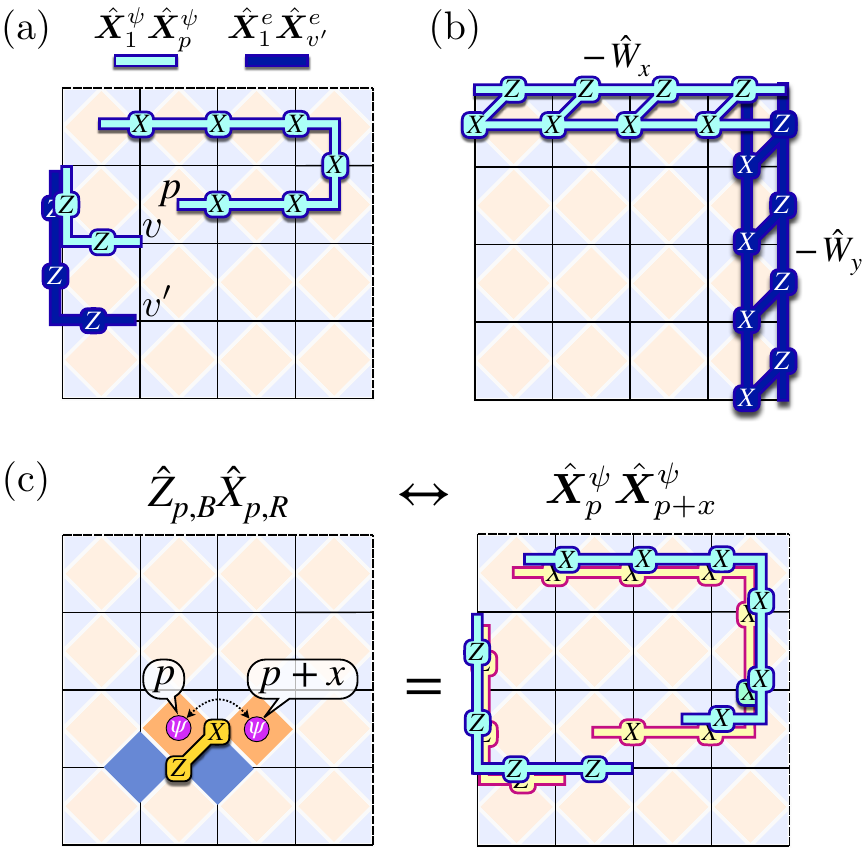}
\caption{(a) Strings of spin operators mapping to the two-body pseudospin operators $\hat{\bm{X}}_1^\psi\hat{\bm{X}}_p^\psi$ and $\hat{\bm{X}}_1^e\hat{\bm{X}}_v^e$. (b) Strings of spin operators realizing the Wilson loop operators of Eq.~\eqref{eq:wilsonloops}. (c) Sketch of the mapping from the two-spin terms $\hat{Z}_{p,B}\hat{X}_{p,R}$, inducing horizontal tunnelling and pairing of fermions on plaquettes $p$ and $p+x$, to two-body pseudospin operators $\hat{\bm{X}}_p^\psi\hat{\bm{X}}_{p+x}^\psi$.}
\label{fig:s2}
\end{figure}

The first step of the quasiparticle mapping is to map the toric-code spins to a dual pseudospin space, such that the commuting operators $\hat{A}_v\hat{B}_{p(v)}$ and $\hat{B}_p$ with $\pm 1$ eigenvalues are mapped to commuting spin-1/2 Pauli-$\hat{Z}$ operators $\hat{\bm{Z}}_v^e$ and $\hat{\bm{Z}}_p^\psi$ living on the corresponding vertex and plaquette. The dual pseudospin space is the tensor product $\mathbb{H}_e \otimes \mathbb{H}_\psi$ of a space $\mathbb{H}_e$ for boson parities and of a space $\mathbb{H}_\psi$ for fermion parities. The pseudospin mapping starts with
\begin{equation} \label{eq:bmzspin}
   \hat{A}_v \hat{B}_{p(v)}\leftrightarrow \hat{\bm{Z}}_v^e, \qquad  \hat{B}_p  \leftrightarrow\hat{\bm{Z}}_p^\psi,
\end{equation}
such that the toric-code Hamiltonian is mapped to
\begin{equation} \label{eq:Htcpseudospin}
\hat{H}_{\rm tc} \leftrightarrow -\frac{g}{2}\sum_v \hat{\bm{Z}}_v^e\hat{\bm{Z}}_{p(v)}^\psi -\frac{g}{2}\sum_p \hat{\bm{Z}}_p^\psi\ .
\end{equation}
On the torus, given the additional constraint 
\begin{equation}
\prod_v \hat{A}_v\hat{B}_{p(v)} = \mathds{1}, \quad \prod_p \hat{B}_{p} = \mathds{1}, 
\end{equation}
the total $\psi$ and $e$ parities are conserved and the local parities do not form a complete set of operators. A complete set is obtained by introducing Wilson-loop operators which commute with all local parities, such as 
\begin{equation} \label{eq:wilsonloops}
\hat{W}_x = -\prod_{p \in \tilde\Gamma_x} {\hat{X}_{p,L} \hat{Z}_{p,T}}, \quad \hat{W}_y = -\prod_{p\in\tilde\Gamma_y}{\hat{X}_{p,B} \hat{Z}_{p,R}},
\end{equation}
where the paths $\tilde\Gamma_{r}$, with $r=x,y$, are depicted in Fig.~\ref{fig:s1} and the operators $\hat{W}_x$ and $\hat{W}_y$ are depicted in Fig.~\ref{fig:s2}(b) [when convenient, we use the notation for spin operators defined after Eq.~(2) and in Fig.~1(a)]. The Wilson-loop operators describe $\psi$-fermion tunnelling all across the torus in horizontal and vertical direction. Their eigenvalues $\pm 1$ distinguish four superselection sectors and, at the end of the mapping, will correspond to the choice of either periodic ($W_x=1$ or $W_y=1$) or anti-periodic ($W_x=-1$ or $W_y=-1$) boundary conditions for the fermions. The operators $\hat{W}_x$ are mapped to two pseudospins $\hat{\bm{Z}}_{r}^W$ in the two-spin space $\mathbb{H}_W$, such that the total pseudospin Hilbert space on the torus has the structure $\mathbb{H}_e\otimes\mathbb{H}_\psi\otimes\mathbb{H}_W$. 
Due to the total fermion and boson parity conservation, single-pseudospin Pauli-$\hat{X}$-type operators which anti-commute with a single $\hat{\bm{Z}}_v^e$ or $\hat{\bm{Z}}_p^\psi$ are not physical and have no mapping in terms of operators on the original spin system: physical spin operators may only create and annihilate anyon pairs, since they always anticommute with an even number of local parities. However, one can define two-body pseudospin operators, anticommuting with $\hat{\bm{Z}}_v^e$ or $\hat{\bm{Z}}_p^\psi$, where one of the two pseudospins is conventionally chosen to be located on the first vertex or plaquette on the top-left of the lattice, which we label as $v=1$ and $p=1$, respectively [Fig.~\ref{fig:s1}(a) and Fig.~\ref{fig:s2}(a)]. These can be defined, for instance, as 
\begin{subequations}\label{eq:xx}
\begin{align}
& \hat{\bm{X}}_1^e  \hat{\bm{X}}_v^e \leftrightarrow \prod_{i\in \Gamma_{1,v}} \hat{Z}_i, \\
& \hat{\bm{X}}_1^\psi  \hat{\bm{X}}_p^\psi \leftrightarrow \prod_{i\in \Gamma_{1,v}}\hat{Z}_i  \prod_{j\in \tilde{\Gamma}_{1,p}}\hat{X}_j, 
\end{align}
\end{subequations}
where $\Gamma_{1,v}$ is a path connecting vertices 1 and $v$, while $\tilde{\Gamma}_{1,p}$ connects the $1$st and the $p$th plaquette, with the specific path convention depicted in Fig.~\ref{fig:s1}(a): $\Gamma_{1,v}$ descends vertically from the top-left vertex until the vertex row to which $v$ belongs and then continues horizontally until $v$; $\tilde{\Gamma}_{1,p}$ connects the top-left plaquette to the right-most one in the same row, descends vertically until the plaquette row to which $p$ belongs, and then horizontally to reach $p$. With this choice, it can be verified that spin anti-commutation rules are satisfied by the pseudospins, $\{\hat{\bm{X}}_1^e \hat{\bm{X}}_v^e,  \hat{\bm{Z}}_v^e\}=0$ and $\{\hat{\bm{X}}_{1}^\psi\hat{\bm{X}}_p^\psi,  \hat{\bm{Z}}_p^\psi\}=0$. The operators in Eq.~\eqref{eq:xx} are depicted in Fig.~\ref{fig:s2}(a).

\begin{figure*}[t]
\centering
\includegraphics[width=\linewidth]{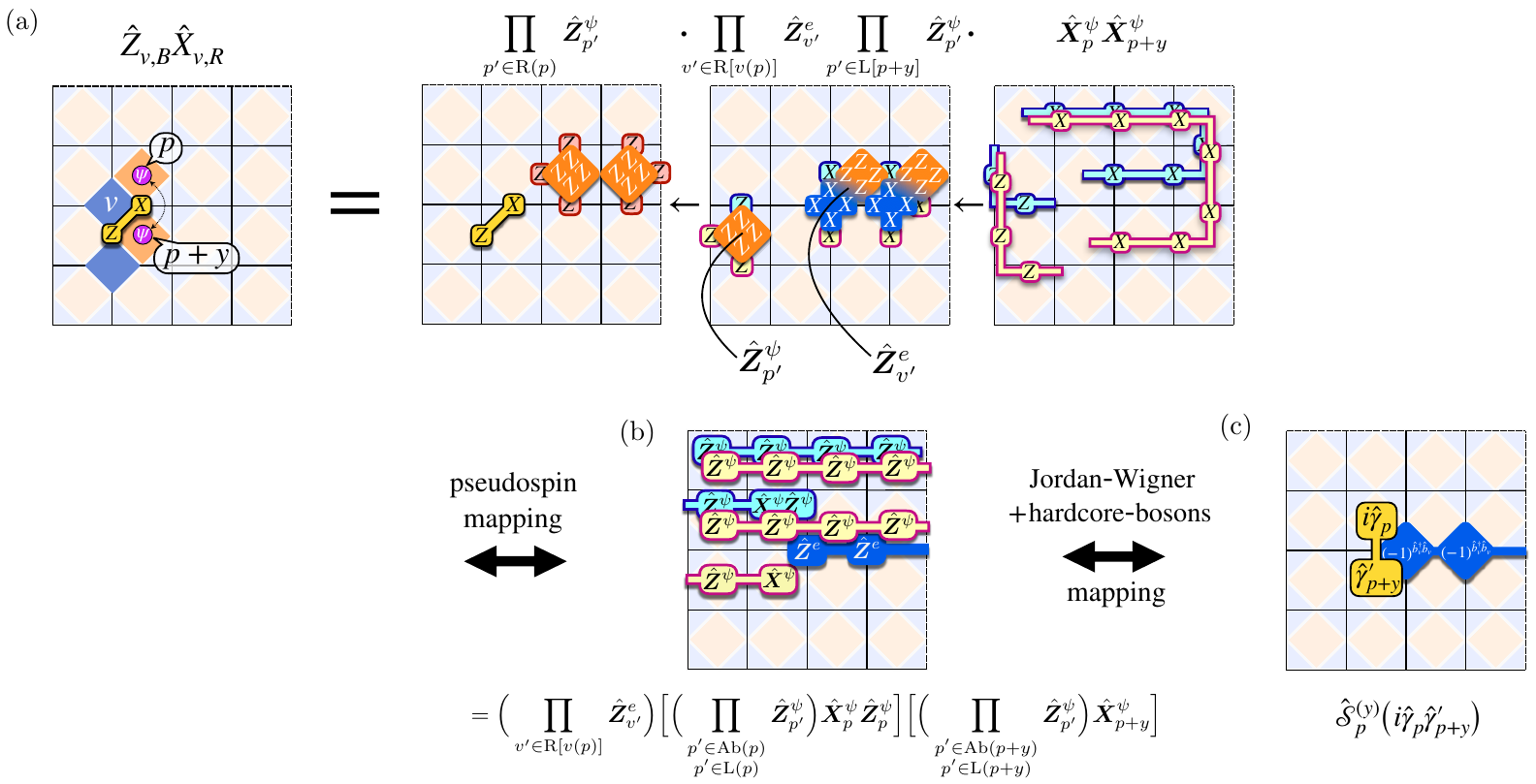}
\caption{Sketch of the different stages of the quasiparticle mapping for vertical fermion processes in the bulk [Eqs.~\eqref{eq:mapp2} and \eqref{eq:verticalproc}]. A two-spin term $\hat{Z}_{v,B}\hat{X}_{v, R}$ inducing tunnelling and pairing of the toric-code fermionic quasiparticles on plaquettes $p$ and $p+y$ is first mapped to a product of pseudospin operators $\hat{\bm{Z}}_p^\psi$, $\hat{\bm{X}}_p^\psi\hat{\bm{X}}_{p'}^\psi$ and $\hat{\bm{Z}}_v^e$ (b). Then, such products are mapped to products of hardcore-boson operators on vertices ($\hat{b}_v$ and $\hat{b}_v^\dagger$, describing $e$ bosons) and Majorana-fermion operators on plaquettes ($\hat{\gamma}_p$ and $\hat\gamma_p'$, related to fermionic operators $\hat{f}_p$ and $\hat{f}_p^\dagger$ for $\psi$ fermions), via a Jordan-Wigner mapping (c).}
\label{fig:s3}
\end{figure*}

Using Eqs.~\eqref{eq:bmzspin} and~\eqref{eq:xx}, the two-spin driving terms of the form $\hat{X}_i\hat{Z}_j$  entering the drive Hamiltonian $\hat{H}_d(t)$ of Eq.~(2) are mapped to pseudospin operators. The precise mapping depends on whether a term involves spins located on the boundaries, which are intersected by $\Gamma_{r}$ or $\tilde\Gamma_{r}$, or not~\cite{Chen2022}. We only report, as an example, the mapping for terms in the bulk, not intersecting the boundary, which reads
\begin{subequations}
\begin{align}
 \hat{X}_{p,R}\hat{Z}_{p,B} \leftrightarrow& \hat{\bm{X}}_{p}^\psi \hat{\bm{X}}_{p+x}^\psi \label{eq:mapp1}\\
 \hat{X}_{v(p),R}\hat{Z}_{v(p),B} \leftrightarrow& \Big[\prod_{v'\in{\mathrm{R}[v(p)]}}\hat{\bm{Z}}_{v'}^e\Big]\nonumber \\
&\times \Big[\prod_{\substack{p'\in \mathrm{L}(p+y)\\ p'\in \mathrm{R}(p)}}\hat{\bm{Z}}_{p'}^\psi\Big]\hat{\bm{X}}_{p}^\psi \hat{\bm{X}}_{p+y}^\psi ,\label{eq:mapp2}
\end{align}
\label{eq:mapstep}
\end{subequations}
\noindent where ${\rm R}(v)$ and $\mathrm{L}(v)$ denote all vertices to the right and to the left of $v$, respectively, and similarly for plaquettes. The mapping of Eq.~\eqref{eq:mapp1} for horizontal processes is sketched in Fig.~\ref{fig:s2}(c), while the mapping of Eq.~\eqref{eq:mapp2} for vertical processes is sketched in Fig.~\ref{fig:s3}(a)-(b).

After the intermediate pseudospin mapping, the quasiparticle mapping is completed by mapping $e$ pseudospins to hardcore bosons and $\psi$ pseudospins to fermions. Specifically, pseudospins of $e$ type are mapped to hardcore bosons via
\begin{equation}
\hat{\bm{Z}}_v^e \leftrightarrow (-1)^{\hat{b}_v^\dagger \hat{b}_v}, \quad \hat{\bm{X}}_v^e \leftrightarrow \hat{b}_v^\dagger + \hat{b}_v,
\end{equation}
where $\hat{b}_v^\dagger$ and $\hat{b}_v$ are bosonic creation and annihilation operators. Equivalently, $\hat{\bm{Z}}_v^e= 1-2\hat{b}_v^\dagger \hat{b}_v$, given the hard-core constraint.
Pseudospins of $\psi$ type are mapped, instead, to fermions via a Jordan-Wigner transformation, 
\begin{equation} \label{eq:jwmap}
\hat{\bm{Z}}_p^\psi \leftrightarrow -i\hat{\gamma}_p \hat{\gamma}_p',  \quad\left[\prod_{p'\in \mathrm{JW}(p)} \hat{\bm{Z}}_{p'}^\psi\right]  \hat{\bm{X}}_p^\psi \leftrightarrow\hat{\gamma}_p',
\end{equation}
where JW$(p)$ is a Jordan-Wigner path with the convention shown in Fig.~\ref{fig:s1}(b) [and exemplified in Fig.~\ref{fig:s3}(b)-(c) for vertical processes in the bulk]: the summation involves all plaquettes scanned left-to-right, row-by-row from the top-left corner of the lattice down, until $p$, thus following the path JW$(p)$ depicted in~\ref{fig:s1}(b). In Eq.~\eqref{eq:jwmap}, we have introduced the Majorana real-fermion operators 
\begin{equation} \label{eq:majorana_gamma}
\hat{\gamma}_p = \hat{f}_p + \hat{f}_p^\dagger, \quad \hat{\gamma}_p' = -i(\hat{f}_p - \hat{f}_p^\dagger),
\end{equation}
which are Hermitian and satisfy $\hat \gamma_p^2=(\hat{\gamma}_p')^2=1$ and, for $p\ne p'$, $\{\hat\gamma_p, \hat\gamma_{p'}\}= \{\hat\gamma_p', \hat\gamma_{p'}'\}= \{\hat\gamma_p, \hat\gamma_{p'}'\}=0$. The operators $\hat{f}_p^\dagger$ and $\hat{f}_p$ are creation and annihilation operators for a complex fermion (representing $\psi$). From Eq.~\eqref{eq:Htcpseudospin}, the toric-code Hamiltonian is then mapped to
\begin{equation} \label{eq:Htfbosongamma}
\hat{H}_{\rm tc} \leftrightarrow \frac{g}{2} \sum_{p}\left[1 + (-1)^{\hat{b}_{v(p)}^\dagger \hat{b}_{v(p)}}\right] i \gamma_p\gamma_p'.
\end{equation}
In terms of complex-fermion operators $\hat{f}_p$ and $\hat{f}_p^\dagger$, Eq.~\eqref{eq:Htfbosongamma} can be expressed as
\begin{multline} \label{eq:Htfbosonfoper}
\hat{H}_{\rm tc} \leftrightarrow 2g\sum_{p}\hat{f}_p^\dagger\hat{f}_p + g\sum_v\hat{b}_v^\dagger\hat{b}_v \\
-2g \sum_p  \hat{b}_{v(p)}^\dagger \hat{b}_{v(p)}\hat{f}_p^\dagger\hat{f}_p,
\end{multline}
as reported in the main text, where we also used the hardcore-boson constraint to express the boson parities as $(-1)^{\hat{b}_v^\dagger \hat{b}_v}=1-2\hat{b}_v^\dagger \hat{b}_v$ and we neglected constant energy shifts.
The two-spin terms of Eq.~\eqref{eq:mapstep} become
\begin{subequations}
\begin{align}
& \hat{X}_{p,R}\hat{Z}_{p,B} \leftrightarrow i \hat{\gamma}_{p} \hat{\gamma}'_{p+x} \label{eq:horizproc} \\
& \hat{X}_{v(p),R}\hat{Z}_{v(p),B} \leftrightarrow \left[\prod_{v\in{\mathrm{R}[v(p)]}}(-1)^{\hat{b}_v^\dagger \hat{b}_v}\right] i \hat{\gamma}_{p} \hat{\gamma}'_{p+y} . \label{eq:verticalproc}
\end{align}
\end{subequations}
The different stages of the mapping leading to Eq.~\eqref{eq:verticalproc} are sketched in Fig.~\ref{fig:s3}. For two-spin terms inducing fermion motion across the boundaries horizontally and vertically, respectively, the mapping yields
\begin{subequations}
\begin{equation}
 \hat{X}_{p,R}\hat{Z}_{p,B} \leftrightarrow \left[\prod_{\substack{ v' \in {\rm Ab}(v)}}(-1)^{\hat{b}_{v'}^\dagger \hat{b}_{v'}}\right] i \hat{\gamma}_{p} \hat{\gamma}'_{p+x} \hat{\bm{Z}}_x^W, 
 \end{equation}
 \begin{equation}
\hat{X}_{v(p),R}\hat{Z}_{v(p),B} \leftrightarrow \left[\prod_{v'\in{\mathrm{R}[v(p)]}}(-1)^{\hat{b}_{v'}^\dagger \hat{b}_{v'}}\right] i \hat{\gamma}_{p} \hat{\gamma}'_{p+y} \hat{\bm{Z}}_y^W,
\end{equation}
\label{eq:boundaryterms}
\end{subequations}
involving the `Wilson-loop pseudospins' $\hat{\bm{Z}}_r^W$, and where Ab$(v)$ indicates all vertices in the vertex rows above the row to which $v$ belongs. In the absence of terms in the Hamiltonian that can induce a change of superselection sector (through tunnelling of a $e$-boson on a full non-contractible loop around the torus), the choice of sector determined by the eigenvalues of the Wilson loop operators $\hat{W}_{x}$ and $\hat{W}_{y}$ then implies the choice of either periodic ($\bm{Z}_x^W=+1$, $\bm{Z}_y^W=+1$) of anti-periodic ($\bm{Z}_x^W=-1$, $\bm{Z}_y^W=-1$) boundary condition for the fermions along each direction. 
Defining the string of boson parities
\begin{equation}
\hat{\mathcal{S}}_{p}^{(y)}=
\left[\prod_{v'\in{\mathrm{R}[v(p)]}}(-1)^{\hat{b}_{v'}^\dagger \hat{b}_{v'}}\right],
\end{equation}
 the driven toric-code Hamiltonian $\hat{H}(t) = \hat{H}_{\rm tc} + \hat{H}_d(t)$, with $\hat{H}_{\rm tc}$ of Eq.~(1) and $\hat{H}_d(t)$ of Eq.~(2), is then mapped, in the bulk, to
\begin{align}
\hat{H}(t) = &  \sum_p \Big\{\frac{g}{2} \left[ 1 + (-1)^{\hat{b}_{v(p)}^\dagger \hat{b}_{v(p)}}\right]i \hat\gamma_p \hat\gamma_{p}' \nonumber \\
& - d_x(t)  i \hat\gamma_p \hat\gamma_{p+x}' - d_y(t) \hat{\mathcal{S}}_p^{(y)}  \ i \hat\gamma_{p} \hat\gamma_{p+y}' \Big\}. \label{eq:mappedwgamma}
\end{align}
Eq.~\eqref{eq:mappedwgamma} applies to bulk fermions: Processes straddling the boundaries are included by including also the terms given in Eqs.~\eqref{eq:boundaryterms}, where the choice of ${\bm{Z}}_x^W$ and ${\bm{Z}}_y^W$ translates into the choice of boundary condition. Note that, since $\hat{S}_p^{(y)}$ is a product of local boson parities and thus commutes with all other bosonic terms entering Eq.~\eqref{eq:mappedwgamma}, the drive does not alter the total boson number $\sum_v\hat{b}_v^\dagger\hat{b}_v$ and does not induce boson tunnelling: a boson Fock state $\ket{\vec{n}^e}=(\hat{b}^\dagger_{v_1})^{n_1}\dots (\hat{b}_{v_N}^\dagger)^{n_N}\ket{0_{\rm tc}}$, listing the $e$ occupations at each vertex for a given distribution of $e$ quasiparticles, is an eigenstate of the bosonic part. This allows us to study the fermion dynamics corresponding to a static background distribution of $e$ quasiparticles. We next make the assumption that, whenever an $e$-boson pair is present at vertices $v$ and $v'$, such that $\bra{\vec{n}^e}\hat{b}_v^\dagger \hat{b}_v\ket{\vec{n}^e}=\bra{\vec{n}^e}\hat{b}_{v'}^\dagger \hat{b}_{v'}\ket{\vec{n}^e}=1$, this corresponds to using a slightly modified toric-code Hamiltonian $\hat{H}_{\mathrm{tc}}'$ in Eq.~(1), with $g\hat{A}_v\to-g\hat{A}_v$ and $g\hat{A}_{v'}\to-g\hat{A}_{v'}$, namely
\begin{equation}
\hat{H}_{\mathrm{tc}}' =\frac{g}{2}(\hat{A}_v+\hat{A}_{v'}) - \frac{g}{2} \sum_{v''\ne v,v' }\hat{A}_{v''}\ - \frac{g}{2} \sum_{p} \hat{B}_{p} ,
\end{equation}
such that the state containing the $e$ pair is the ground state with respect to the modified Hamiltonian. While this has no effect on the driving terms $\hat{X}_i\hat{Z}_j$ and their quasiparticle mapping, it changes the free toric-code part of Eqs.~\eqref{eq:Htfbosongamma} and \eqref{eq:Htfbosonfoper}, which now
maps to 
\begin{multline}
\hat{H}_{\rm tc}' \leftrightarrow \frac{g}{2} \sum_{v''\ne v,v'}\left[1 + (-1)^{\hat{b}_{v''}^\dagger \hat{b}_{v''}}\right] i \gamma_{p(v'')}\gamma_{p(v'')}' \\
\frac{g}{2} \sum_{v''\in\{v, v'\}}\left[1 - (-1)^{\hat{b}_{v(p)}^\dagger \hat{b}_{v(p)}}\right] i \gamma_p\gamma_p' ,
\end{multline}
such that its expectation value with respect to the boson state $\ket{\vec{n}^e}$, containing the $e$-boson pair at vertices $v$ and $v'$, gives
\begin{equation}
\bra{\vec{n}^e}\hat{H}_{\rm tc}'\ket{\vec{n}^e} = g\sum_p i \gamma_p\gamma_p'\ .
\end{equation} 
This assumption is not important for the non-Abelian physics studied but it slightly simplifies the analysis of the fermion Hamiltonian, which then has a homogeneous chemical potential for all $p$ also in the presence of $e$ quasiparticles, thus directly mapping to the simplest common models of spinless superconductors~\cite{Sachdev2023}. 

Under this assumption and including the drive, the Hamiltonian~\eqref{eq:mappedwgamma}, for a fixed configuration of $e$ bosons defined by a boson Fock state $\ket{\vec{n}^e}$, gives the bulk fermion Hamiltonian $\hat{H}_\psi(t)  = \bra{\vec{n}^e} \hat{H}_{\rm tc} + \hat{H}_d(t) \ket{\vec{n}^e}$ reading as
\begin{multline}
\hat{H}_\psi(t)  = \sum_p \Big\{ g i\hat\gamma_p \hat\gamma_{p}'  - d_x(t)  i \hat\gamma_p \hat\gamma_{p+x}'  \\
- d_y(t) \hat{\mathcal{S}}_p^{(y)}  \ i \hat\gamma_{p} \hat\gamma_{p+y}' \Big\} .
\end{multline}
Re-expressing the Majorana operators in terms of complex fermions $\hat{f}_p$ and $\hat{f}_p$ via Eq.~\eqref{eq:majorana_gamma}, defining the fermion chemical potential $\mu_\psi=2g$ and neglecting constant energy shifts then gives the fermion Hamiltonian
\begin{multline}
\hat{H}_\psi(t)  = \mu_\psi \sum_p \hat{f}_p^\dagger \hat{f}_p  \\
- d_x(t)  \sum_p (\hat{f}_p\hat{f}_{p+x} + \hat{f}_p^\dagger \hat{f}_{p+x}+ \mathrm{H.c.}) \\
- d_y(t)\sum_p \hat{\mathcal{S}}_p^{(y)} (\hat{f}_p\hat{f}_{p+y} + \hat{f}_p^\dagger \hat{f}_{p+y}+ \mathrm{H.c.})  .
\end{multline}
 corresponding to Eq.~(3) of the main text.

Consider now $e$-boson (vortex) movement. This is induced by the single-spin operator $\hat{Z}_i$. In the quasiparticle mapping, for spins in the bulk, this is mapped to~\cite{Chen2022} 
\begin{align}
&\hat{Z}_{v,R} \leftrightarrow (\hat{b}_{v}^\dagger + \hat{b}_{v})(\hat{b}_{v+x}^\dagger + \hat{b}_{v+x}), \nonumber\\
& \hat{Z}_{v,T} \leftrightarrow \hat{\tilde{\mathcal{S}}}_v^{(y)}(\hat{b}_{v}^\dagger + \hat{b}_{v})(\hat{b}_{v+y}^\dagger + \hat{b}_{v+y}), \label{eq:tunne}
\end{align}
involving a string of fermion operators
\begin{equation}
\label{eq:Stilde}
\hat{\tilde{\mathcal{S}}}_v^{(y)}= \left[\prod_{p \in \mathrm{L}[p(v+y)]} (-i\hat{\gamma}_p\hat{\gamma}_{p}') \right].
\end{equation}
Hence, horizontal tunnelling has `no strings attached', while vertical vortex tunnelling carries a string of fermion parities spanning over all the plaquettes to the left of the bond connecting vertices $v$ and $v+y$.

\section{Floquet engineering}

\subsection{Effective Hamiltonian}

In this section, the high-frequency Floquet Hamiltonian of Eq.~(5) is derived. The latter is produced by the driven fermion Hamiltonian $\hat{H}_{\psi}(t)=\hat{H}_{\psi}(t+T)$ of Eq.~(3), which we report here,
\begin{multline}
\hat{H}_\psi(t)  =   \sum_{p} \Big[\mu_{\psi} \hat{f}_{p}^\dagger \hat{f}_{p}\\- \!\! \sum_{r\in\{x, y\}}  d_{r}(t) {\mathcal{S}}_{p}^{(r)}  \big(\hat{f}_{p} \hat{f}_{p+r} 
+ \hat{f}_{p}^\dagger \hat{f}_{p+r} + \mathrm{H. c.}\big) \Big]. 
\label{eq:reportedH}
\end{multline}  
Following Floquet's theorem, the evolution operator generated by a time-periodic Hamiltonian $\hat{H}(t)=\hat{H}(t+T)$, starting from an initial time $t_0$, can be decomposed according to~\cite{Shirley1965, Goldman2014, Bukov2015,Eckardt2017}
\begin{equation} \label{eq:floquetdecomp}
\hat{U}(t,t_0) = \hat{U}_F(t,t_0) e^{-i \hat{H}_{F}^{[t_0]} (t-t_0)}
\end{equation}
in terms of a time-independent Floquet Hamiltonian $\hat{H}_{F}^{[t_0]}$ and a time-periodic micromotion operator $\hat{U}_F(t,t_0)=\hat{U}_F(t+T, t_0)=e^{\hat{K}(t,t_0)}$, generated by the `kick' operator $\hat{K}(t,t_0)$. The stroboscopic dynamics in steps of $T$ is then fully ruled by  $\hat{H}_{F}^{[t_0]}$ according to
\begin{equation}
\hat{U}(t_0+nT,t_0) = e^{-inT \hat{H}_F^{[t_0]}}.
\end{equation}
 The superscript in $\hat{H}_{F}^{[t_0]}$ makes explicit the parametric dependence on the initial time $t_0$ of the evolution and thus on the global phase of the drives at that time. As detailed in the following, for the driving scheme analyzed here, the stroboscopic generators at different $t_0$ (as well as the non-stroboscopic Floquet Hamiltonian~\cite{Eckardt2015}) match to leading order in high frequency and exhibit the desired topological properties. For this reason, in the main text we dropped the superscript on the Floquet Hamiltonian and denoted it with $\hat{H}_{F}$ (the impact of the phase of the drives on the non-Abelian vortex exchange phase is quantitatively analyzed below in Sec.~\ref{sec:numericaldet}). Both $\hat{H}_{F}^{[t_0]}$ and $\hat{K}(t,t_0)$ can be determined order by order in a high-frequency expansion in powers of $\omega^{-1}$~\cite{Eckardt2015, Bukov2015}. The leading order is given by
\begin{subequations}
\begin{align}
& \hat{H}_F^{[t_0]} = \hat{H}_0= \frac{1}{T}\int_{0}^{T} \hat{H} (t) dt, \label{eq:H0} \\
& \hat{K}(t,t_0) = -\sum_{m\ne 0} \frac{\left(e^{im\omega t}-e^{im\omega t_0}\right)\hat{H}_m }{m\omega} , \label{eq:G}
\end{align}
\end{subequations}
where we have indicated the Fourier components of $\hat{H}(t)$ with
\begin{equation}
\hat{H}_m = \frac{1}{T}\int_0^T \! \!dt\  \hat{H}(t) e^{-im\omega t}\ .
\end{equation}
To compute the expansion for the Floquet Hamiltonian produced by $\hat{H}_\psi(t)$ of Eq.~\eqref{eq:reportedH}, it is convenient to first represent $\hat{H}_{\psi}(t)$ in a rotating frame defined by the gauge transformation
\begin{align} \label{eq:Rtransf}
\hat{R}(t) = \exp\left(-i t \frac{\omega}{2}\sum_{p}\hat{f}_{p}^\dagger\hat{f}_{p}\right),
\end{align}
such that the evolution operator decomposes as $\hat U(t,t_0)=\hat R(t) \hat V(t, t_0)$. 
At the end of one period, $\hat R(T)$ reduces to the total fermion parity of the system, $\hat{R}(T) = \exp\left( -i\pi \sum_{p}\hat{f}_{p}^\dagger \hat{f}_{p}\right)=(-1)^{\sum_p \hat{f}_p^\dagger \hat{f}_p}$, which is a conserved quantity at all times, $[\hat{R}(T), \hat{H}_\psi(t)]=0$. Moreover, $\hat R(T)$ reduces to the identity within the physical subspace of even-parity states, and thus $\hat{R}(t)$ is $T$-periodic in that subspace.
The transformed Hamiltonian $\hat{H}'_{\psi}(t) = \hat{R}^\dagger(t) \hat{H}_\psi(t)\hat{R}(t) -i \hat{R}^\dagger(t) \partial_t \hat{R}(t)$, generating $\hat V(t,t_0)$, is also $T$-periodic and reads
\begin{align}
& \hat{H}'_\psi(t)  = \sum_{p} \Big[  (\mu_{\psi} - \omega/2)\hat{f}_{p}^\dagger\hat{f}_{p}\nonumber \\
& - \!\! \sum_{r\in\{x, y\}}  d_{r}(t) \hat{\mathcal{S}}_{p}^{(r)}  \big(e^{i \omega t}\hat{f}_{p} \hat{f}_{p+r} + \hat{f}_{p}^\dagger \hat{f}_{p+r} + \mathrm{H. c.}\big) \Big].
\label{eq:Hprimet}
\end{align}
Applying Floquet's theorem to $\hat{H}'_\psi(t)$, the propagator $\hat V(t,t_0)$ can be decomposed as in Eq.~\eqref{eq:floquetdecomp} with a Floquet Hamiltonian $\hat{H}_F^{[t_0]\prime}$. 
Inserting the driving modulation $d_{r}(t)=J + 2\Delta\cos(\omega t+\phi_r)$ of Eq.~(4) into Eq.~\eqref{eq:Hprimet}, the Hamiltonian $\hat{H}'_\psi(t)$ possesses the following time dependence,
\begin{widetext}
\begin{multline}
 \hat{H}'_\psi(t)  = \sum_{p} \Big\{  (\mu_{\psi} - \omega/2)\hat{f}_{p}^\dagger\hat{f}_{p}
 - \!\! \sum_{r\in\{x, y\}}   \hat{\mathcal{S}}_{p}^{(r)}  J\big(e^{i \omega t}\hat{f}_{p} \hat{f}_{p+r} + \hat{f}_{p}^\dagger \hat{f}_{p+r} + \mathrm{H. c.}\big)\\
 - \sum_{r\in\{x, y\}}  \hat{\mathcal{S}}_{p}^{(r)}\Delta  \big[e^{i\phi_{r}}\big(\hat{f}_{p} \hat{f}_{p+r} + e^{i\omega t} (\hat{f}_{p}^\dagger \hat{f}_{p+r} + \hat{f}_{p+r}^\dagger \hat{f}_{p}) 
+ e^{i2\omega t}\hat{f}_{p} \hat{f}_{p+r}\big)+ \mathrm{H. c.}\big]\Big\} .\label{eq:Heffstep}
\end{multline}
\end{widetext}
The leading order of the high-frequency expansion for the Floquet Hamiltonian, $\hat{H}_F^{[t_0]\prime}$ is obtained by time averaging [Eq.~\eqref{eq:H0}], which yields the time-independent part of Eq.~\eqref{eq:Heffstep}, namely
\begin{multline}
\hat{H}_{\rm avg}^{[t_0]\prime}= \sum_{p} \Big\{-\mu\hat{f}_{p}^\dagger\hat{f}_{p}
 - \!\! \sum_{r\in\{x, y\}}  \hat{\mathcal{S}}_{p}^{(r)}  \big( J \hat{f}_{p}^\dagger \hat{f}_{p+r} \\ + \Delta e^{i\phi_{r}} \hat{f}_{p} \hat{f}_{p+r} + \mathrm{H. c.}\big)\Big\} ,%
 \label{eq:almostHF}
\end{multline}
where we have defined the effective chemical potential $\mu=\omega/2-\mu_\psi$. This corresponds to the Hamiltonian of Eq.~(4) of the main text. In the original reference frame [before transforming through $\hat R(T)$], the approximated Floquet Hamiltonian $\hat{H}_{\rm avg}^{[t_0]}$ is determined from the relation 
\begin{align}
e^{-i \hat{H}_{\rm avg}^{[t_0]} T} &= \hat{U}(t_0+T,t_0) = \hat{R}(T)\hat V(t_0+T,t_0) \nonumber \\
& = e^{-i\pi\sum_p\hat{f}_p^\dagger \hat{f}_p} e^{-i  \hat{H}_{\rm avg}^{[t_0]\prime}T}
\label{eq:fromprimeto}
\end{align}
and the fact that $[\hat{R}(T), \hat{H}_{\rm avg}^{[t_0]\prime}]=0$ due to the conservation of the total fermion parity. The effect of $\hat{R}(T)$ is thus simply to shift the zero Bogoliubov--de Gennes quasienergy by $\pi/T$, that is half a Floquet-Brillouin zone. 
 By choosing the gauge $\hat{f}_p\to e^{-i\phi_x/2}\hat{f}_p $, it becomes clear that the effective Hamiltonian of Eq.~\eqref{eq:almostHF} only depends, to leading order in high-frequency, on the phase-delay $\phi_y - 
\phi_x$ between the horizontal and vertical drives, and not on the global phase of the drives (the  impact of the latter is analyzed in the following Sec.~\ref{sec:numericaldet}).
The kick operator $\hat{K}(t,t_0)$, to leading order, is determined by Eq.~\eqref{eq:G}, given $\hat{R}(t)$ and the Fourier components 
of $\hat{H}'_\psi(t)$. The latter, from Eq.~\eqref{eq:Heffstep}, read
\begin{align}
\hat{H}' _1  =&  - \sum_{p}\sum_{r\in\{x, y\}}   \hat{\mathcal{S}}_{p}^{(r)}J \hat{f}_{p} \hat{f}_{p+r} \nonumber \\
& - \sum_{p}\sum_{r\in\{x, y\}}  \hat{\mathcal{S}}_{p}^{(r)} \Delta e^{i\phi_{r}} (\hat{f}_{p}^\dagger \hat{f}_{p+r} + \hat{f}_{p+r}^\dagger \hat{f}_{p}) ,
\end{align}
\begin{align}
\hat{H}' _{-1}  =&  - \sum_{p}\sum_{r\in\{x, y\}}   \hat{\mathcal{S}}_{p}^{(r)}   J \hat{f}_{p+r}^\dagger \hat{f}_{p}^\dagger \nonumber \\
& - \sum_{p}\sum_{r\in\{x, y\}}   \hat{\mathcal{S}}_{p}^{(r)}\Delta e^{-i\phi_{r}}(\hat{f}_{p}^\dagger \hat{f}_{p+r} + \hat{f}_{p+r}^\dagger \hat{f}_{p}) ,
\end{align}
\begin{align}
 \hat{H}'_2  =&-\sum_{p}  \sum_{r\in\{x, y\}} \hat{\mathcal{S}}_{p}^{(r)}   \Delta  e^{i\phi_{r}}\hat{f}_{p} \hat{f}_{p+r} ,
\end{align}
\begin{align}
 \hat{H}'_{-2}  =&-\sum_{p}  \sum_{r\in\{x, y\}} \hat{\mathcal{S}}_{p}^{(r)}   \Delta  e^{-i\phi_{r}} \hat{f}_{p+r}^\dagger \hat{f}_{p}^\dagger .
\end{align}

\subsection{Numerical determination of the Floquet Hamiltonian}
\label{sec:numericaldet}

In order to determine the effective Floquet Hamiltonian numerically, the Hamiltonian $\hat{H}_\psi(t)$ of Eq.~\eqref{eq:reportedH} is first represented in Bogoliubov-de Gennes (BdG) form,
\begin{equation}
\hat H_\psi(t) = (1/2)(\hat{\bm{f}}^\dagger, \hat{\bm{f}}) \hat H_{\rm BdG}(t) (\hat{\bm{f}}, \hat{\bm{f}}^\dagger)^T,
\end{equation}
where $\hat{\bm{f}}=\{\hat{f}_1, \ldots, \hat{f}_N\}$ and $\hat{\bm{f}}^\dagger=\{\hat{f}_1^\dagger, \ldots, \hat{f}_N^\dagger\}$, with $N=L\times L$, which involves a time-periodic BdG Hamiltonian $\hat H_{\rm BdG}(t)$. The effective Floquet-BdG Hamiltonian $\hat{H}_{F,\mathrm{BdG}}$ generated by $\hat H_{\rm BdG}(t)$ according to Floquet's decomposition [Eq.~\eqref{eq:floquetdecomp}] is determined numerically by diagonalizing the end-of-period propagator $\hat{U}_{\rm BdG}(t_0+T,t_0)$ determined through direct time propagation of the Schr\"odinger Eq. 

\begin{figure}
 \includegraphics[width=0.8\linewidth]{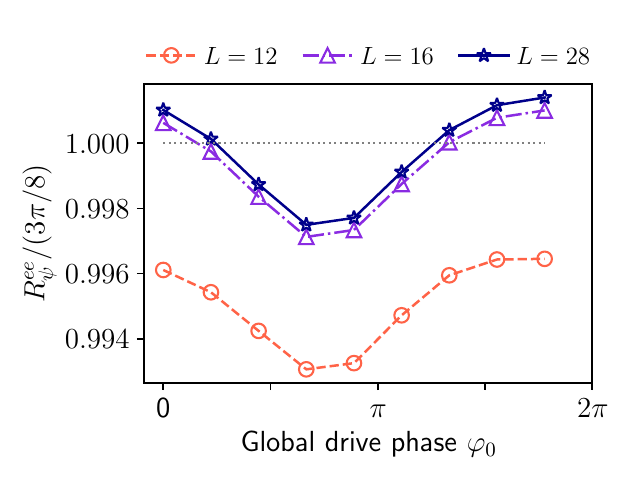}
 \caption{Vortex exchange phase $R_{\psi}^{ee}$ in the $\psi$ sector, divided by its ideal value $3\pi/8$, as a function of the global phase $\varphi_0$ of the driving functions, defined by $d_x(t)=J+2\Delta\cos(\omega t + \varphi_0)$ and $d_y(t)=J+2\Delta\cos(\omega t + \varphi_0 + \pi/2)$. The relative phase $\pi/2$ imprints $p+ip$ pairing, as discussed in the main text. The driving amplitude and frequency are set at $J=\Delta=\mu_\psi/16$ and $\omega=2\mu_\psi-4J=7\mu_\psi/4$ in the topological region. The gray dotted line is a guide for the eye at the ideal value 1.}
 \label{fig:s4}
 \end{figure}

 As discussed in the main text and detailed in Sec.~\ref{sec:appendix_bdg}, the calculation of the exchange phases builds on the assumption that ground states of the effective Floquet Hamiltonian for the driven fermions, corresponding to a given distribution of $e$ quasiparticles, are prepared. In the high-frequency regime considered here, all $\hat{H}_{F,\mathrm{BdG}}^{[t_0]}$ for any initial time $t_0$ coincide to leading order in $1/\omega$ [see Eq.~\eqref{eq:G}]: the Hamiltonians are well approximated by the time-averaged Hamiltonian of Eq.~\eqref{eq:almostHF}, which possesses entirely the desired topological properties for the effective fermion bands, and only differ by small corrections due to the weak higher-order terms. This is indeed verified by comparing the vortex exchange phases obtained using ground states of $\hat{H}_{\rm F,BdG}^{[t_0]}$ for different initial time $t_0$, as shown in Fig.~\ref{fig:s4}. From Fig.~\ref{fig:s4}, one can appreciate that, for all system sizes shown, the variation in the exchange phase is within $\sim0.3\%$.
 
\section{Vortex exchange phases}
\label{sec:appendix_bdg}

In this section, we discuss how the ground state of the Floquet Hamiltonian is constructed and used to compute geometric phases when $e$ vortices are displaced.\\

\textbf{Bogoliubov-de Gennes quasiparticles.} \label{sec:BdG}
 The fermion Floquet Hamiltonian takes the general form
\begin{equation}
\hat{H} = -\sum_j \mu_j n_j - \sum_{ i j} \big(J_{ij}\hat f_i^\dagger \hat f_j + \Delta^*_{ij} \hat f_i \hat f_j + \mathrm{H.c.}\big)
\end{equation}
and can be expressed as $\hat H = (1/2)(\hat{\bm{f}}^\dagger, \hat{\bm{f}}) \hat H_{\rm BdG} (\hat{\bm{f}}, \hat{\bm{f}}^\dagger)^T$
in terms of the Bogoliubov-de Gennes Hamiltonian $\hat H_{\rm BdG}$. Diagonalizing the BdG Hamiltonian, 
\begin{align}
& \hat{H}_{\rm BdG} \begin{pmatrix}
\bm{u}_n \\
\bm{v}_n
\end{pmatrix} = E_n \begin{pmatrix}
\bm{u}_n \\
\bm{v}_n
\end{pmatrix}\\
& \hat{H}_{\rm BdG} \begin{pmatrix}
\mp\bm{v}_n^* \\
\bm{u}_n^*
\end{pmatrix} = -E_n \begin{pmatrix}
\mp\bm{v}_n^* \\
\bm{u}_n^*
\end{pmatrix}, \label{eq:negeneig}
\end{align}
gives pairs of particle-hole symmetric energies $\pm E_n$ and eigenvectors. One can use them to construct quasiparticle operators $\hat{\alpha}_n$,
\begin{equation}
\begin{pmatrix}
\hat\alpha_{1},\\
\vdots \\
\hat\alpha_{N},\\
\hat\alpha_{1}^\dagger,\\
\vdots \\
\hat\alpha_{N}^\dagger\\
\end{pmatrix} =
\begin{bmatrix}
\bm{u}_{1}^{*,T}, \bm{v}^{*,T}_{1} \\
\vdots \\
\bm{u}_{N}^{*,T}, \bm{v}^{*,T}_{N} \\
\bm{v}_{1}^T, \bm{u}^T_{1} \\
\vdots \\
\bm{v}_{N}^T, \bm{u}^T_{N} \\
\end{bmatrix}
\begin{pmatrix}
\hat{f}_{1},\\
\vdots \\
\hat{f}_{N},\\
\hat{f}_{1}^\dagger,\\
\vdots \\
\hat{f}_{N}^\dagger\\
\end{pmatrix} 
= \hat{\mathcal{U}}^\dagger \begin{pmatrix}\bm{f}\\ \bm{f}^\dagger \end{pmatrix},
\end{equation}
where $\hat{\mathcal{U}}$ is of the form
\begin{equation} \label{eq:Uform}
\hat{\mathcal{U}} = \begin{pmatrix}
u& v^* \\
v & u^*
\end{pmatrix},
\end{equation}
 such that $\hat{H}$ is brought to canonical form
\begin{equation}
\hat{H} = E_{\rm GS} + \sum_{n=1}^N E_n \hat\alpha_n^\dagger \hat\alpha_n ,
\end{equation}
with ground state energy
\begin{equation}
E_{\rm GS} = -\frac{1}{2}\sum_{n=1}^N E_n.
\end{equation}
The ground state $\ket{\Phi}$ is determined by the condition of being annihilated by all quasiparticle operators, $\hat{\alpha}_n\ket{\Phi}=0$. It can be written explicitly in generalized BCS form~\cite{RingSchuck1980},
\begin{equation} \label{eq:Phi0}
\ket{\Phi} = \mathcal{N}\exp\left(\frac{1}{2}\sum_{ij} T_{ij} \hat{f}_i^\dagger \hat{f}_j^\dagger\right)\ket{0_{\rm tc}} ,
\end{equation}
with a Thouless matrix $T_{ij}$, with respect to the chosen reference state $\ket{0_{\rm tc}}$, which is the fermion vacuum. The normalization constant is $\mathcal{N} ={\mathrm{det} (u)}$, with $u$ defined in Eq.~\eqref{eq:Uform}. In case $\ket{\Phi}$ has vanishing overlap with the fermion vacuum, $\langle \Phi |0_{\rm tc}\rangle=0$, a different reference state needs be chosen, as further discussed below. By imposing $\hat\alpha_n\ket{\Phi}=0$, the Thouless matrix $T_{ij}$ is found as $T = -(u^*)^{-1} v^*$~\cite{RingSchuck1980}.
Since, for the toric-code model on a torus, only pairs of fermionic quasiparticles can be created or annihilated, the physical states must have even fermionic parity. The ground state parity is determined by the Floquet-BdG Hamiltonian and is computed following Ref.~\cite{Kitaev2006}. Namely, the Floquet Hamiltonian is first expressed in terms of Majorana operators [Eq.~\eqref{eq:majorana_gamma}] as $\hat{H}_F= i\sum_{p,p'} {H}^{(\gamma)}_{F, pp'}\hat \gamma_p\hat\gamma_{p'}'$. The ground state parity is then given by the Pfaffian $\mathrm{Pf}( B)$ of the real skew-symmetric matrix $ B$ defined as
\begin{equation}
B =  Q\begin{pmatrix}
0 & 1 &  & &\\
-1 & 0 & & &\\
&  & \ddots & &  \\
& & &0 & 1\\
& & &-1 & 0
\end{pmatrix} Q^T,
\end{equation}
where $Q$ is the matrix decomposing ${H}^{(\gamma)}_F$ in the form 
\begin{equation}
{H}^{(\gamma)}_F =  Q\begin{pmatrix}
0 & \varepsilon_1 &  & &\\
-\varepsilon_1 & 0 & & & \\
&  & \ddots &  &\\
& & & 0 & \varepsilon_N\\
& & & -\varepsilon_N & 0
\end{pmatrix}  Q^T.
\end{equation}
In case the lowest energy state of $\hat{H}_{\rm BdG}$ has odd parity, the physical ground state will be the one containing the first BdG quasiparticle. This can be determined as the state which is annihilated by all the operators $\alpha_j$ with $j\ne 1$ and by $\alpha_1^\dagger$ \cite{RingSchuck1980, Chen2022}.\\

\textbf{Geometric phase.} 
To compute the geometric phase accumulated by moving a vortex along a given path, we follow the strategy of Refs.~\cite{Chen2022, Levin2003}, adapted here to the Floquet system. 
We consider controlled vortex movement within the bulk, that never spans the whole lattice, such that no flipping of the eigenvalues of the Wilson loop operators (change of superselection sector) can occur. For each position of the vortices along the path, the fermion Floquet-BCS corresponding to that vortex configuration is computed from the BdG Hamiltonian as explained in the previous section (see also Sec.~\ref{sec:numericaldet}), resulting in a sequence of fermion ground states $\ket{\Phi_j}$. To avoid dependence of the phase from the details of the exchange path, Levin-Wen's protocol~\cite{Levin2003} is employed, which requires one to compute the difference $\theta=\theta_{P} -\theta_{P'} $ of the phases $\theta_{P}$ accumulated along two `T-shaped' paths depicted in Fig.~3(a). The phase along path $P$ is computed as
\begin{equation} \label{eq:phasepath}
\theta_{P} = \mathrm{Arg}\prod_{(i,i+1) \in P} \bra{\Phi_{i+1}} \hat{t}_{i+1, i} \ket{\Phi_i}.
\end{equation}
Here, the operator $\hat{t}_{i,j}$ represents the fermionic part of the $e$-boson tunnelling operators of Eq.~\eqref{eq:tunne} which converts the vortex spatial configuration $\ket{\vec{n}_j^e}$, associated with the fermion BCS ground state $\ket{\Phi_j}$, into $\ket{\vec{n}_i^e}$, associated with $\ket{\Phi_i}$. In the quasiparticle picture, $\hat{t}_{i,j}$ is given by the string $\hat{\tilde{\mathcal{S}}}_p^{(y)}$ of fermion parities of Eq.~\eqref{eq:Stilde}. Hence, 
the matrix elements in Eq.~\eqref{eq:phasepath} reduce, for horizontal vortex displacement, to computing the overlap between `neighbouring' BCS wavefunctions, while, for vertical displacement, one needs to compute matrix elements of the form
\begin{equation} \label{eq:matrix_el}
\bra{\Phi_{i+1}} \hat{\tilde{\mathcal{S}}}_p^{(y)}\ket{\Phi_i}.
\end{equation}

To compute overlaps, we choose $\ket{\Phi_1}$ as a reference state to construct all other Floquet BCS ground states $\ket{\Phi_\ell}$, rather than the bare fermion vacuum as in Eq.~\eqref{eq:Phi0}, according to
\begin{equation}
\label{eq:phiell}
\ket{\Phi_\ell} = \mathcal{N}_\ell \exp\left(\frac{1}{2}\sum_{i,j} T^{(\ell)}_{ij} \alpha_{\ell,i}^\dagger \alpha_{\ell,j}^\dagger\right) \ket{\Phi_1}.
\end{equation}
On the one hand, this choice of reference state `saves' one computation step for $\theta_P$ by fixing the phase in the first step to $\mathrm{Arg}\langle\Phi_2|\Phi_1\rangle=0$, since the ansatz of Eq.~\eqref{eq:phiell} implicitly fixes the gauge along the path through the condition that $\langle \Phi_i|\Phi_1\rangle=\mathcal{N}_1$ is real. On the other, it is a convenient choice since $\langle \Phi_i|\Phi_1\rangle$ typically remains fairly large, avoiding numerical issues in the construction of the state associated with small overlaps. 

The Thouless matrix $T_{ij}^{(\ell)}$ and operators $\hat{\bm{\alpha}}_\ell$ can be determined by using the eigenvectors of the 1st and $\ell$th Floquet-BdG Hamiltonians, which give rise to the two sets of quasiparticle operators
\begin{align}
\begin{pmatrix}
\hat{\bm{\alpha}}_1 \\
\hat{\bm\alpha}_1^\dagger 
\end{pmatrix}  = \hat{\mathcal{U}}_1^\dagger \begin{pmatrix}
\hat{\bm{f}} \\
\hat{\bm{f}}^\dagger 
\end{pmatrix} , \quad
\begin{pmatrix}
\hat{\bm{\alpha}}_\ell \\
\hat{\bm{\alpha}}_\ell^\dagger 
\end{pmatrix}  = \hat{\mathcal{U}}_\ell^\dagger \begin{pmatrix}
\hat{\bm{f}} \\
\hat{\bm{f}}^\dagger 
\end{pmatrix},
\end{align}
respectively, where $\hat{\mathcal{U}}_\ell$ have the form of Eq.~\eqref{eq:Uform} in terms of submatrices $u_\ell$ and $v_\ell$.
The matrix $T^{(\ell)}$ is then given by $T^{(\ell)}=-(\tilde{u}_\ell^*)^{-1} \tilde{v}_\ell^{*}$, where the matrices $\tilde{u}_\ell$ and $\tilde{v}_\ell$ are defined by~\cite{RingSchuck1980}
\begin{subequations}
\label{eq:uandv}
\begin{align}
 \tilde{u}_\ell =&  u_1^\dagger u_\ell + v_1^\dagger v_\ell, \\
  \tilde{v}_\ell =& v_1^T u_\ell + u_1^T v_\ell. 
\end{align}
\end{subequations}

Given two ground states $\ket{\Phi_i}$ and $\ket{\Phi_j}$, characterized by Thouless matrices $T^{(i)}$ and $T^{(j)}$, their overlap can be computed according to~\cite{Robledo2009}
\begin{equation}
\langle\Phi_i|\Phi_j\rangle = (-1)^{N(N+1)/2} \mathrm{pf}
\begin{pmatrix}
T^{(j)} & -\mathds{1} \\
\mathds{1} & -T^{(i),*}
 \end{pmatrix} ,
 \label{eq:overlap}
\end{equation}
where $\mathrm{pf}(\cdot)$ denotes the Pfaffian.
To compute the matrix elements of Eq.~\eqref{eq:matrix_el}, we note that, if $\ket{\Phi}$ is the quasiparticle vacuum associated with the BdG Hamiltonian $\hat H_{\rm BdG}$, then $\ket{\Phi'}=\hat Q \ket{\Phi}$ with $\hat Q$ unitary is a quasiparticle vacuum associated with the transformed BdG Hamiltonian 
\begin{equation}
\hat H_{\rm BdG}' = \hat Q \hat H_{\rm BdG} \hat{Q}^\dagger\ .
\end{equation}
Given the matrix $\hat U$ of eigenvectors of $\hat H_{\rm BdG}$, the matrix $\hat U'$ of eigenvectors of $\hat H_{\rm BdG}'$ is then $\hat U' = \hat Q \hat U$. The latter can then be used to construct the representation of the BCS vacuum $\ket{\Phi'}$ with respect to the reference state $\ket{\Phi_1}$, according to Eqs.~\eqref{eq:phiell} and~\eqref{eq:uandv}, and then be used to compute its overlap with $\ket{\Phi}$ via Eq.~\eqref{eq:overlap}. However, there is an additional caveat. Using the representation~\eqref{eq:phiell} fixes the gauge of $\ket{\Phi'}$ by imposing a real $\langle{\Phi'| \Phi_1}\rangle$, which is unwanted. Indeed, consistency with the gauge choice made along the path, $\langle{\Phi_i| \Phi_1}\rangle\in \mathbb{R}$, only requests that each ground state $\ket{\Phi}$ (and not also $\hat Q\ket{\Phi}$) has real overlap with the reference state $\ket{\Phi_1}$. To remedy for this, in the computation of the phase in Eq.~\eqref{eq:matrix_el} we introduce a fictitious intermediate step to $\ket{\Phi_i'}$ to fix the gauge choice, namely
\begin{equation} \label{eq:intstep}
\bra{\Phi_j} \hat Q \ket{\Phi_i} \longrightarrow \langle\Phi_j|\Phi_i'\rangle \langle \Phi_i'| \Phi_i\rangle.
\end{equation}
To see that this re-establishes the correct phase, note that $\ket{\Phi'}$ (in the BCS representation enforcing a real $\langle{\Phi'| \Phi_1}\rangle$) is equal to $\hat{Q}\ket{\Phi}$ up to some phase $\theta$, $\ket{\Phi'}=e^{i\theta}\hat{Q} \ket{\Phi}$. Inserting the latter expression into Eq.~\eqref{eq:intstep}, one obtains that the phase $\theta$ cancels out, leaving
 \begin{equation} \label{eq:anstep}
 \langle\Phi_j|\Phi_i'\rangle \langle \Phi_i'| \Phi_i\rangle =  \langle\Phi_j | \hat Q | \Phi_i\rangle \langle \Phi_i | \hat Q |\Phi_i\rangle.
\end{equation}
If $\hat Q$ is Hermitian, as $\hat{\tilde{\mathcal{S}}}_p^{(y)}$ of Eq.~\eqref{eq:matrix_el} is, the diagonal element $\langle \Phi_i| \hat{Q}|\Phi_i\rangle$ is real, and thus the phase contribution in Eq.~\eqref{eq:anstep} only comes from $\langle\Phi_j | \hat Q | \Phi_i\rangle$ as desired.

 \section{Driving-induced heating}
 
Periodically-driven quantum systems are generically expected to heat up to a featureless infinite-temperature-like state at long times as a consequence of energy absorption from the drive~\cite{Dalessio2014, Lazarides2014, Moessner2017}. In this section, we analyze these aspects for the driven model studied in this work, elaborating on the statements given in the main text. First of all, since the local state space (spins) is bounded, one expects that the timescale over which heating kicks in can be prolonged exponentially with increasing driving frequency~\cite{Abanin2015, Mori2016}. Moreover, a key observation is that our driven toric code maps to a model of static $e$ boson and driven \textit{noninteracting} fermions in the quasiparticle picture. Such integrable systems are predicted not to heat up to a featureless state, due to the existence of an extensive number of conserved quantities given by the occupations of the single-particle Floquet modes~\cite{Russomanno2012, Lazarides2014b}, although the expectation value of certain observables may still resemble the infinite-temperature case~\cite{Ishii2018}. In the following, we present numerical results confirming that no heating takes place in the high-frequency regime of interest, whereas a strong energy increase is observed away from it when vortices are present.

\subsection{Numerical analysis}
\label{sec:heating2}

To investigate the deviation of the exact Floquet dynamics from the time averaged description, resulting in heating, we will inspect heating measures for varying ratio $J/\omega$ between the driving amplitudes $J=\Delta$ and the driving frequency $\omega$, which is the relevant perturbative parameter in the high-frequency expansion. Recall that, in the time-averaged model of Eq.~(5), the detuning of the driving frequency $\omega$ from the fermion-pair gap $2\mu_\psi$ determines the effective chemical potential $\mu=\omega/2-\mu_\psi$ for the driven fermions. We analyse a regime in which the time-averaged model is in the topological phase, namely $\mu=-2J$, implying $\omega = 2\mu_\psi - 4J$. Thus, to maintain non-vanishing $\omega$, we will consider the range $J/\omega\le 1$. 

\begin{figure}
\centering
\includegraphics[width=\linewidth]{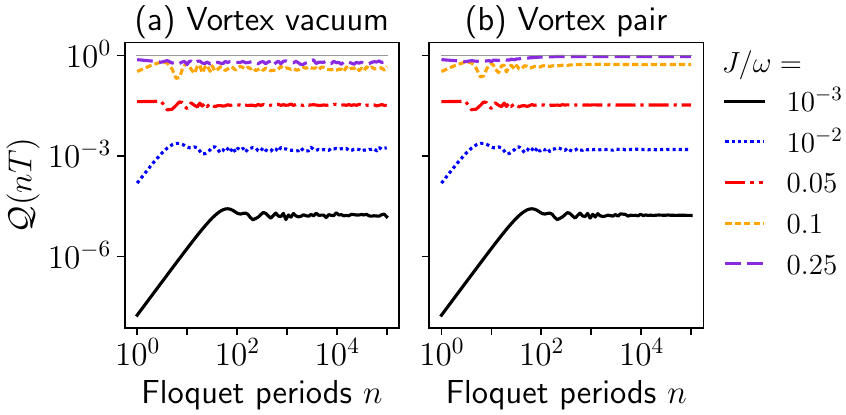}
\caption{Stroboscopic value of $\mathcal{Q}(n T)$ of Eq.~\eqref{eq:QnT} as a function of the number of Floquet period, in the absence (a) and presence (b) of a vortex pair, for antiperiodic boundary conditions for the fermions. The results in (b) are averaged over 50 random $e$-pair configurations. The standard deviation obtained is barely visibile on the scale of the plot, indicating little dependence on the vortex position, and we thus did not include it. The thin gray line is a guide for the eye at value 1.}
\label{fig:QnT}
\end{figure}

To investigate the impact of heating during the time evolution, we first consider the stroboscopic dynamics $\ket{\psi(nT)}$ with the system initially prepared in the fermion ground state $\ket{\Psi_0}$ of the time-averaged Hamiltonian $\hat{H}_{\rm avg}$ of Eq.~(5), for fixed background distributions of $e$ particles. The fact that the exact Floquet Hamiltonian differs from the time-averaged one will lead to deviations in the expectation value of physical observables, which may be considered as heating~\cite{Eckardt2017}. 
 To verify this, we monitor heating through the quantity~\cite{Heyl2019} 
\begin{equation} \label{eq:QnT}
\mathcal{Q}(n T) = \frac{\bra{\psi(nT)} \hat{H}_{\rm avg} \ket{\psi(nT)} - E_{0} }{E_\infty - E_{0}},
\end{equation}
where $E_{0} = \bra{\Psi_0} \hat{H}_{\rm avg} \ket{\Psi_0}$ is the initial and ideal value, and $E_{\infty}=(1/2^N) \mathrm{tr}(\hat{H}_{\rm avg})$ is the infinite-temperature value. Thus, $\mathcal{Q}(nT)=0$ if no heating occurs, whereas $\mathcal{Q}(nT)=1$ if an infinite-temperature-like value is reached. Technical details about the calculation of $\mathcal{Q}(n T)$ are given in Sec.~\ref{sec:measures} in the following. The results for $L=24$ are shown in Fig.~\ref{fig:QnT}, for different ratio $J/\omega$. The panel~\ref{fig:QnT}(a) depicts the result in the absence of vortices, whereas~\ref{fig:QnT}(b) corresponds to the average of 50 random configurations of a single pair of $e$ particles. Recall that the presence of vortices impacts the dynamics by inducing anyonic effects as the fermions move. For ratios in in the high-frequency regime ($J/\omega \ll 1$), $\mathcal{Q}(nT)$ saturates well below the infinite-temperature value both in the presence and in the absence of vortex pairs with similar behaviour. Moreover, as predicted, heating is strongly suppressed with decreasing $J/\omega$. The time-averaged Hamiltonian thus remains a good description of the system even at long times. A different behaviour is observed when $J/\omega \sim 1$, corresponding to the regime where the high-frequency expansion starts to diverge: the value of $\mathcal{Q}(nT)$ is much larger instead and approaches the infinite-temperature value.

\begin{figure}
\includegraphics[width=\linewidth]{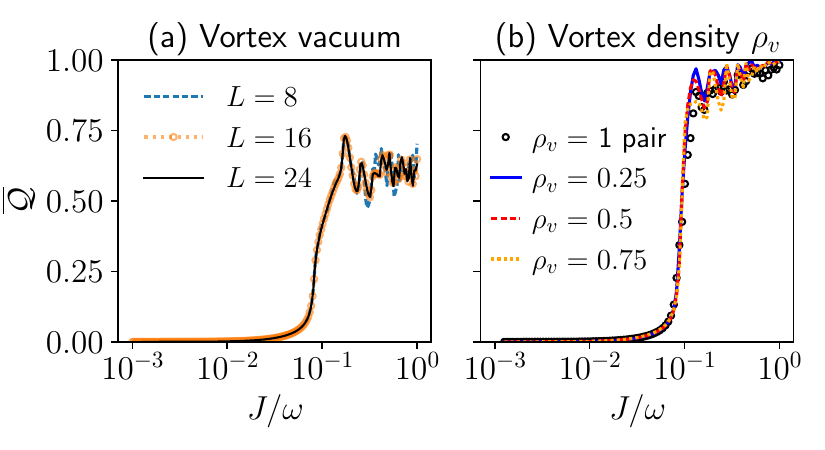}
\caption{Infinite-time average $\overline{\mathcal{Q}}$ of Eq.~\eqref{eq:barQ} as a function of the ratio $J/\omega$ for different system size in the absence of vortices (a) and for $L=16$ and different vortex densities $\rho_v$ (b), with antiperiodic fermion boundary conditions. The results in (b) are averaged over 20 random vortex configurations at the given density (b). The standard deviation obtained is barely visibile on the scale of the plot, indicating little dependence on the vortex position, and we thus did not display it. }
\label{fig:barQ}
\end{figure}

To investigate the long-time dynamics more systematically, we analyze the infinite-time average~\cite{Ishii2018}
\begin{equation} \label{eq:barQ}
\overline{\mathcal{Q}} = \frac{\bra{\Psi_0} \overline{H}_{\rm avg} \ket{\Psi_0} - E_0}{E_\infty - E_0}.
\end{equation}
 The Hamiltonian $\overline{H}_{\mathrm{avg}}$ is obtained from $\hat{H}_{\rm avg}$ by neglecting the off-diagonal matrix elements in the basis of Floquet modes. This is reported in Fig.~\ref{fig:barQ} as a function of the ratio $J/\omega$ for different system size $L=8, 16, 24$. The panel~\ref{fig:barQ}(a) reports the results in the absence of vortices. In~\ref{fig:barQ}(b), we investigate the impact of vortices by considering random distributions of $e$ particles at a given density $\rho_v$ and then averaging $\overline{Q}$ over 20 configurations for each $\rho_v$ and $L=16$. In the high-frequency regime, $\overline{\mathcal{Q}}$ is always close to zero independently from the presence of vortices: the system does not heat up, confirming the practicality of the Floquet engineering scheme proposed. In the anomalous Floquet regime, where the time-averaged description is not valid anymore, $\overline{\mathcal{Q}}$ increases significantly, but still saturates below the infinite-temperature state in the absence of vortices and exhibits little dependence on the system size. The situation is different in the presence of vortices, where values closer to $\overline{\mathcal{Q}}=1$ are found. This indicates that, although the many-body occupation of the single-particle Floquet modes is conserved, the dynamics with vortices strongly mixes the single-particle eigenstates of $\hat{H}_{\rm avg}$, yielding infinite-temperature-like expectation values of the one-body observable $\hat{H}_{\rm avg}$ at long times~\cite{Ishii2018}.

\subsection{Calculation of the heating measures} 
\label{sec:measures}

In this subsection, we explain how the figures of merit $\mathcal{Q}(nT)$ and $\overline{\mathcal{Q}}$ of Eqs.~\eqref{eq:QnT} and~\eqref{eq:barQ} used to quantify heating in Sec.~\ref{sec:heating2} are computed. The key quantities needed are the expectation values
 \begin{subequations}
  \begin{align}
 &\mathrm{(i)} \bra{\Psi_0}\hat{U}(nT)^\dag \hat{H}_{\rm avg} \hat{U}(nT) \ket{\Psi_0}, \label{eq:(ii)2}\\
  &\mathrm{(ii)} \bra{\Psi_0}\overline{{H}}_{\rm avg} \ket{\Psi_0}. \label{eq:(i)1} 
 \end{align}
 \end{subequations}
which are of the general form $\bra{\Phi}\hat{H} \ket{\Phi}$ for a Hamiltonian $\hat{H}$ and a BCS state $\ket{\Phi}$. 
To compute such quantities, consider the BdG representation of $\hat{H}$, $\hat{H}=(1/2)(\hat{\bm{f}}^\dagger, \hat{\bm{f}}) \hat{H}^{\rm BdG} (\hat{\bm{f}}, \hat{\bm{f}}^\dagger)^T$ and the quasiparticle operators $(\hat{\bm{\alpha}}, \hat{\bm{\alpha}}^\dag)^T = \hat{\mathcal{U}}(\hat{\bm{f}}, \hat{\bm{f}}^\dagger)^T$ which annihilate the BCS state $\ket{\Phi}$ (see Sec.~\ref{sec:appendix_bdg}), where $\hat{\mathcal{U}}$ has the form of Eq.~\eqref{eq:Uform}, namely
\vspace{-0.5cm}

\begin{equation} \label{eq:Uformagain}
\hat{\mathcal{U}} = \begin{pmatrix}
u& v^* \\
v & u^*
\end{pmatrix}.
\end{equation}
The expectation value $\langle \hat{H}\rangle = \bra{\Phi}\hat{H} \ket{\Phi}$ then reads as
\begin{widetext}
\begin{align}
\langle \hat{H} \rangle  =  \frac{1}{2}\sum_{i,j=1}^N \hat{H}^{\rm BdG}_{i,j} \langle \hat{f}_i^\dagger \hat{f}_j \rangle +  \frac{1}{2}\sum_{i,j=1}^N \hat{H}^{\rm BdG}_{i,j+N} \langle \hat{f}_i^\dagger \hat{f}_j^\dag \rangle 
  +  \frac{1}{2}\sum_{i,j=1}^N \hat{H}^{\rm BdG}_{i+N,j} \langle \hat{f}_i \hat{f}_j \rangle + \frac{1}{2}\sum_{i,j=1}^N \hat{H}^{\rm BdG}_{i+N, j+N} \langle \hat{f}_i \hat{f}_j^\dag \rangle .
\end{align}
\end{widetext}
Expressing the operators $(\hat{\bm{f}}^\dag, \hat{\bm{f}})$ in terms of $(\hat{\bm{\alpha}}^\dag, \hat{\bm{\alpha}})$ by means of $\hat{\mathcal{U}}$ and Eq.~\eqref{eq:Uformagain}, one then finds
\begin{align}
& \langle \hat{f}_i^\dagger \hat{f}_j \rangle = (v v^\dag)_{ij},&  \langle \hat{f}_i^\dagger \hat{f}_j^\dag \rangle = (u v^\dag)_{ij} , \nonumber \\
& \langle \hat{f}_i \hat{f}_j \rangle = (v u^\dag)_{ij}, &  \langle \hat{f}_i \hat{f}_j^\dag \rangle = (u u^\dag)_{ij}\ .
\end{align}
Returning to the specific expectation values in Eqs.~\eqref{eq:(ii)2} and \eqref{eq:(i)1}, the Hamiltonian $\overline{H}_{\rm avg}$ of~\eqref{eq:(ii)2}  is obtained, as discussed in Sec.~\ref{sec:heating2}, by dropping the off-diagonal elements of $\hat{H}_{\rm avg}$ in the basis of Floquet states. The Hamiltonian $\hat{U}(nT)^\dag \hat{H}_{\rm avg} \hat{U}(nT)$ of \eqref{eq:(ii)2} is computed from the numerically-determined end-of-period propagator $\hat{U}(T)$ and Floquet theorem, implying $\hat{U}(nT) = [\hat{U}(T)]^n$.

\end{document}